\documentclass[twocolumn,aps,prd,showpacs,nofootinbib,superscriptaddress,preprintnumbers,floatfix,10pt]{revtex4-1}

\usepackage{amsmath}
\usepackage{amsfonts}
\usepackage{float} 
\usepackage{amssymb}
\usepackage{epsfig}
\usepackage{graphicx}
\usepackage{euscript}
\usepackage{color}
\usepackage{slashed}
\usepackage{epstopdf}
\usepackage[utf8]{inputenc}
\usepackage[normalem]{ulem}

\usepackage{hyperref}

\newcommand{\be}{\begin{equation}}
\newcommand{\ee}{\end{equation}}
\newcommand{\bea}{\begin{eqnarray}}
\newcommand{\eea}{\end{eqnarray}}

\newcommand{\nn}{\nonumber\\}
\newcommand{\al}[1]{\begin{align}#1\end{align}}
\newcommand{\I}[2]{\mathcal{I}^{#1}_{#2}}
\newcommand{\lgCross}{{g}^0_{4,1}}
\newcommand{\lgCyclicI}{{g}^{2,i}_{4,1}}
\newcommand{\lgCyclic}[1]{{g}^{2,#1}_{4,1}} 

\newcommand{\lgCatI}{{g}^{1,i}_{6,1}}
\newcommand{\lgCat}[1]{{g}^{1,#1}_{6,1}}

\newcommand{\lgCyclicDisI}{{g}^{3,i}_{6,2}}
\newcommand{\lgCyclicSixI}[1]{{g}^{3,i}_{6,1}}

\newcommand{\dg}[4]{{{g}}^{#1,#2}_{#3, #4}}
\newcommand{\dgI}[3]{{{g}}^{#1,i}_{#2, #3}}
\newcommand{\dgC}[3]{{{g}}^{#1}_{#2, #3}}
\newcommand{\Scaling}[3]{[{{\bar{g}}}^{#1,i}_{#2,#3}]}
\newcommand{\ScalingC}[3]{[{{\bar{g}}}^{#1}_{#2,#3}]}

\begin{document}

\preprint{}

\title {Towards background independent quantum gravity with tensor models
}

\author{Astrid Eichhorn}
\email[]{a.eichhorn@thphys.uni-heidelberg.de} 
\affiliation{Institute for Theoretical Physics, University of Heidelberg, Philosophenweg 16, 69120 Heidelberg, Germany}

\author{Johannes Lumma}
\email[]{j.lumma@thphys.uni-heidelberg.de} 
\affiliation{Institute for Theoretical Physics, University of Heidelberg, Philosophenweg 16, 69120 Heidelberg, Germany}

\author{Tim Koslowski}
\email[]{koslowski@nucleares.unam.mx} 
\affiliation{Instituto de Ciencias Nucleares, Universidad Nacional Aut\'onoma de M\'exico, Cto. Exterior S/N, C.U., Del. Coyoac\'an, CDMX, M\'exico}

\author{Antonio D.~Pereira}
\email[]{a.pereira@thphys.uni-heidelberg.de}
\affiliation{Institute for Theoretical Physics, University of Heidelberg, Philosophenweg 16, 69120 Heidelberg, Germany}

\begin{abstract}
 We explore whether the phase diagram of tensor models
 could
 feature a pregeometric, discrete and a geometric, continuum phase for the building blocks of space. The latter are associated to rank $d$ tensors of size $N$.
We search for a universal large $N$ scaling limit in a rank-3 model with real tensors that could be linked to a transition between the two phases. We 
extend
the conceptual development and practical implementation of the flow equation for the pregeometric setting.  
This
provides a 
pregeometric ``coarse-graining"   
by going from many microscopic to few effective degrees of freedom by lowering $N$.  
We discover several candidates for fixed points of this coarse graining procedure, and specifically explore the impact of a novel class of interactions allowed in the real rank-3 model. In particular, we explain how most universality classes feature dimensional reduction, while one candidate, involving a tetrahedral interaction, might potentially be of relevance for three-dimensional quantum gravity.
\end{abstract}


\maketitle
\section{Introduction}

Tensor models \cite{Gurau:2010ba,Gurau:2011aq,Gurau:2011xp,Rivasseau:2011hm,Rivasseau:2012yp,Rivasseau:2013uca,Rivasseau:2016zco,Rivasseau:2016wvy} have been put forward as a generalization of matrix models \cite{DiFrancesco:1993cyw,David:1984tx,David:1985nj,Ambjorn:1985az,Kazakov:1985ea,Boulatov:1986mm,Boulatov:1986jd} to higher dimensions and provide a setting to explore random geometries with a view towards developing a quantum theory of gravity \cite{Ambjorn:1990ge,Sasakura:1990fs,Godfrey:1990dt,Gross:1991hx,Gurau:2010ba,Gurau:2011aq}. The Feynman diagram expansion of tensor models can be mapped to triangulations, i.e., discretized configurations of space, i.e., random geometries. Topological fluctuations are typically included in this setting, with the exception of models which are constructed specifically to suppress these \cite{Benedetti:2008hc}.\\
The first key challenge in this setting is to discover whether a universal continuum limit in these models can be taken. This requires the existence of an appropriate, universal large $N$ limit where $N$ is the size of the tensors. In this limit, the model transitions from finitely to infinitely many degrees of freedom and might potentially encode the dynamics of continuum gravity.\\
The second key challenge is to characterize the physics of the continuum phase in order to settle whether tensor models can provide a description of spacetime compatible with our universe.\\
In this paper, we tackle the first challenge in three-dimensional models as a stepping stone towards the four-dimensional case. 

While computer simulations are making progress towards finding a universal continuum limit in dynamically triangulated discrete quantum gravity \cite{Ambjorn:2000dv,Ambjorn:2001cv,Laiho:2016nlp} with candidates for the required higher-order phase transition found in Causal Dynamical Triangulations, \cite{Ambjorn:2011cg,Ambjorn:2012ij,Ambjorn:2017tnl}, probing the tentative continuum limit in simulations is computationally rather expensive. We instead pursue a complementary way of exploring dynamical triangulations. It is specifically adapted to probe systems at criticality, and is versatile in being applicable to different models. The functional Renormalization Group (FRG), established for these models in \cite{Eichhorn:2013isa,Eichhorn:2014xaa,Eichhorn:2017xhy}, and later also
extended
to group field theories (GFTs), e.g., in \cite{Benedetti:2014qsa,Benedetti:2015yaa,Geloun:2015qfa,Geloun:2016qyb,Lahoche:2016xiq,Carrozza:2016tih,Geloun:2016xep,Carrozza:2016vsq,Carrozza:2017vkz,BenGeloun:2018ekd,Lahoche:2018oeo,Lahoche:2018ggd} as well as to causal sets \cite{Eichhorn:2017bwe} is our tool of choice. 
 See also \cite{Carrozza:2016vsq} for a review of flows for tensorial field theories and \cite{Carrozza:2014rya,Carrozza:2016tih} for an asymptotically safe fixed point in a GFT.
The FRG is closely related to the Polchinski equation for these models \cite{Krajewski:2015clk,Krajewski:2016svb}, and provides a nonperturbative tool complemented by Dyson-Schwinger equations for related models \cite{Sanchez:2017gxt,Pascalie:2018nrs}. 

Besides their interest for quantum gravity, tensor models play a role in aspects of condensed-matter models, see, e.g., \cite{1206.5539v2,Witten:2016iux,Klebanov:2016xxf,Gurau:2016lzk,Klebanov:2018fzb}  and have recently also been applied to study the emergence of turbulence in resonant systems \cite{Evnin2018}. While the purely combinatorical models of interest in the quantum-gravity context often get supplemented by a background space or spacetime, the combinatorical structure of interest in condensed matter is the same or similar. Accordingly, our work also provides some information on the combinatorial aspects of universality classes in these models. 

In \cite{Eichhorn:2017xhy} the FRG equation was applied to rank-3 complex tensor models to find a possible universal continuum limit of these models. Universality is key in this context, as it implies independence from microscopic details. In the setting of dynamical triangulations, the microscopic building blocks, e.g., tetrahedra, are introduced as a discretization of spacetime in order to regularize the gravitational path integral \footnote{Note that arguments, based, e.g., on black-hole entropy, that quantum gravity should be discrete, do \emph{not} preclude that one should take the continuum limit. In fact, as for instance the example of discrete spectra of geometric operators in Loop Quantum Gravity \cite{Rovelli:1994ge,Ashtekar:1996eg} together with the ongoing search for a continuum limit, see, e.g., \cite{Dittrich:2014ala} highlight, discreteness can be a dynamically emergent property. See also \cite{Eichhorn:2017bwe,Eichhorn:2018yfc} for further discussions of this question.}. Accordingly, no physical meaning should be ascribed to the choice of building blocks, and universality is required in the continuum limit in order to reach independence of emergent large-scale properties from these unphysical choices.
Candidates for a universal continuum limit can be found as fixed points of the FRG equation:
\bea
  \partial_t \Gamma_N[T]& =& N\, \partial_N\, \Gamma_N[T] \nonumber\\
  &=& \frac{1}{2}{\rm Tr}\left[\left(\Gamma_N^{(2)}[T]+R_N \right)^{-1}\partial_t R_N\right],\label{equ:FRGE}
\eea
 for the flowing action $\Gamma_N[T]$.
$R_N$ denotes an infrared (IR)-suppression 
term that gives a ``mass" (i.e., a quadratic term) $\Delta_N S[T] = \frac{Z_N}{2} T_{abc}R_N^{abcdef}T_{def}$ (with the Einstein summation convention)
of order $N$ to the tensor elements $T_{abc}$ with index values $a+b+c<N$. Here $Z_N$ stands for the wave-function renormalization.  The IR is associated to small index values and the ultraviolet (UV) to large index values. This corresponds to an intuitive understanding of UV in the sense of many degrees of freedom (large tensors) and IR as fewer degrees of freedom (small tensors), where the effect of the UV modes is encoded in the effective dynamics $\Gamma_N$ for the effective degrees of freedom. This notion of a pregeometric RG flow that does not rely on notions of local coarse graining but simply interpolates between many degrees of freedom and few degrees of freedom is in line with the a-theorem \cite{Cardy:1988cwa}.
Fixed points satisfying $\partial_t\Gamma_N=0$ provide universality classes that are candidates for a limit to a physical continuum quantum gravity regime. We stress that while a fixed point  in this abstract RG flow does not come with the usual notion of scale-invariance, its crucial consequence is that of universality  as advocated in 
\cite{Rivasseau:2011hm,Rivasseau:2012yp,Rivasseau:2013uca,Rivasseau:2016wvy}.
 
In the present work we go beyond \cite{Eichhorn:2017xhy} by not only exploring the real model \cite{Carrozza:2015adg} which features an additional, potentially key interaction  first introduced by Carrozza and Tanasa \cite{Carrozza:2015adg}, but also illuminating the notion of scaling dimension in these pregeometric models in more detail. 
Our work paves the way for a study of rank-4 tensor models which could play a key role in the description of quantum spacetime in our universe.\\
The remainder of the paper is organized as follows: 
In section~\ref{Theory space} we explore the technical aspects of the theory space spanned by tensor models. In particular, we discuss how the RG flow in a pregeometric setting differs
from the one in standard quantum field theory. 
Accordingly, many notions familiar from quantum field theories on a background, such as, e.g., scaling dimensions, have to be re-investigated. We discuss the conceptual basics for this and then show how to do this in practice.
In section~\ref{sec:model}, we introduce the model that we are studying in this work and explain how to set up the RG flow. In section~\ref{sec:universalityclasses} we search for universality classes. That is, we explain how we distinguish physical fixed points from unphysical, truncation-induced zeros of the beta functions
and establish a clear set of necessary criteria that a physical fixed point has to satisfy. In section~\ref{uniclassdimred} we present  several
potential universality classes that we argue exhibit dimensional reduction. We find three candidate universality classes belonging to that type. In section~\ref{qgcandidate} we present a candidate for a universality class for 3d quantum gravity. The fixed point 
underlying this
universality classes is a novel feature of the real model compared to the complex one. Finally in section~\ref{sec:conclusions} we conclude and provide an outlook towards applications of our results to quantum gravity.

\section{Theory space}\label{Theory space}
Background independence is a key requirement of quantum-gravity models. It extends the general relativistic insight that there is no fixed background, and spacetime is dynamical, to the quantum setting. Yet, background independence in a quantum setting comes at a conceptual as well as technical ``cost", as most quantum-field-theoretic tools are based on the presence of a background. For instance, tensor models differ conceptually from field theories on a background spacetime in that they do not possess a canonical notion of locality, dimension and scale separation. These are essential ingredients for the Renormalization Group (RG) in its standard implementation as a local coarse graining. This warrants some conceptual considerations before delving into a concrete RG setup. 

\subsection{The RG flow and the infinite cutoff limit in a pregeometric setting}
 Given a path integral defined with an appropriate cutoff, such that it is finite, one can transform the problem of performing the integral into the problem of solving a (functional) differential equation \cite{Wetterich:1992yh}. This is done by introducing a cutoff term that suppresses a subset of configurations in the path integral as a function of an adjustable parameter. If that cutoff term is chosen to be quadratic in the fields, a relatively simple differential equation follows that encodes the response of the effective action to a change in that parameter. In the continuum, these steps can be implemented in the form of a local coarse-graining procedure. Conceptually, this underlies the Wilsonian RG setup, where a family of scale-dependent effective dynamics encodes the same physics in a way adapted to the scale. For clarity, we sketch the derivation in the standard setting here: Starting with a path integral
 \be
 Z[J] = \int \mathcal{D}\varphi \, e^{-S[\varphi]+ \int J\cdot \varphi},
 \ee
 a cutoff term $R_k(p^2)$ is introduced that acts as a mass-like suppression for modes with ``wavelength" $1/p>1/k$,
 \be
 Z_k[J] = \int \mathcal{D}\varphi\, e^{-S[\varphi]+ \int J\cdot \varphi- \frac{1}{2}\int_p \varphi(p)R_k(p^2)\varphi(-p)},\label{eq:Zk}
 \ee
such that $Z_{k\rightarrow 0}[J]= Z[J]$. Instead of the scale-derivative $k\partial_k\, Z_k$, one now focuses on
  the scale derivative of the so-called effective average action
  \be
\Gamma_k[\phi]= \underset{J}{\rm sup} \left( \int J \cdot \phi - {\rm ln}Z_k[J]\right)- \frac{1}{2} \int_p \phi(p) R_k(p^2)\phi(-p),
\ee
with 
\be
\langle \varphi \rangle_k = \phi.
\ee
Taking the scale derivative $k\partial_k$, one can directly derive the equation
\be
k\partial_k\, \Gamma_k = \frac{1}{2}\int_p \left(\frac{\delta^2\Gamma_k}{\delta\phi(p)\delta\phi(-p)}+R_k(p^2) \right)^{-1}k\partial_k\, R_k(p^2),\label{eq:floweqcont}
\ee
where the second functional derivative of $\Gamma_k$ appears because the  
regulator term in Eq.~\eqref{eq:Zk} is quadratic in the field. These steps have transformed the original task of performing a functional integral into the task of solving a functional differential equation. Irrespective of the concepts underpinning this equation, the formal manipulations allowing the transformation between an integral and a differential equation can be done in any setting.\\
In the background-dependent setting, the flow equation Eq.~\eqref{eq:floweqcont} implements a Wilsonian integration over successive momentum shells, corresponding to a local form of coarse-graining.
 In the pregeometric case, the formal manipulations leading to a ``flow equation" are straightforward to implement
 and lead to Eq.~\eqref{equ:FRGE}. Yet, it is not immediately obvious that there is an underlying physical interpretation. As already discussed in \cite{Eichhorn:2014xaa}, a physical interpretation can be given if the parameter with respect to which configurations are suppressed is chosen as the tensor size $N$, see \cite{Brezin:1992yc}. This gives meaning to coarse-graining in a pregeometric setting. Specifically, large $N$ corresponds to the limit with many degrees of freedom, i.e., the UV. Conversely, small $N$ corresponds to the IR, where most quantum fluctuations have been integrated out and only a few, effective degrees of freedom remain. Phrased in this way, the conceptual similarity between a local, geometric, and a pregeometric coarse-graining become obvious. In particular, the pregeometric flow respects the intuition that leads to the a-theorem for RG flows \cite{Cardy:1988cwa}.\\
The setup and interpretation of a flow equation as a means to construct the full effective action is only one aspect that we are interested in here. The other central question pertains to the construction of the infinite-cutoff limit. In the geometric setting, since the coarse graining is local, one can use the flow equation ``towards the UV" to find out whether a continuum limit can be taken. Since a continuum limit is tied to a scale-invariant regime, a local form of coarse graining is well-suited to asking this question. In the pregeometric setting, the analogous question is more intricate. Here, we are interested in taking the limit of infinitely many degrees of freedom, i.e., $N \rightarrow \infty$. This limit can be taken in a self-similar regime, in which appropriately rescaled couplings no longer depend on $N$. While self-similarity is  the feature that is in common between the geometric and the pregeometric setting, its interpretation differs: In the geometric case, it is an actual self-similarity under a process of ``zooming in'' locally in spacetime. In the pregeometric case, it is a self-similarity of the dynamics for large and small tensors. This requires that appropriately rescaled couplings stay finite as $N \rightarrow \infty$. The  rescaling factors $N^{d_{\bar{g}_i}}$ are the analogues of the canonical scaling dimensions, and are determined by requiring that the beta functions have a well-defined large $N$ expansion and become autonomous in the large-$N$ limit.

In the following, we proceed to elaborate on the second of the above aspects, and discuss the existence of a universal large-$N$ scaling limit in more formal terms.

Functional RG equations are a very versatile (non-perturbative) tool for the investigation of path integrals, which in turn are a framework to perform the highly non-trivial task of constructing measures $d\mu(T)$ with given symmetry requirements on infinite dimensional spaces with fields $T$. 
In the present paper we consider real tensors $T_{abc}$ where the index values range over all positive integers.
A measure on an infinite dimensional space can be encoded in a partition function 
$Z[J]:=\int d\mu(T)\exp(
{\rm Tr (J\cdot T)}
)$, where  
$J$
is a source dual to the field $T$. Measures $d\mu(T)$ on an infinite dimensional space restrict to measures on finite dimensional spaces $d\mu_{N^\prime}(T)$. This is implemented by a coarse-graining procedure. With the restricted measure, we can evaluate the expectation value $\int d\mu(T)\,f(T)$ of a function $f(T)$ that depends only on a finite number of components of the tensor $T_{abc}$. It is convenient to choose these finite dimensional spaces such that the functions $f_{N^\prime}(T)$ on these finite dimensional spaces depend only on tensor components whose index values satisfy $a+b+c\le N^\prime$. (In the continuum, the analogous requirement is the insistence on a local coarse graining, instead of a different way of ``sifting" field configurations in the path integral.)
For any such function $f_{N^\prime}(T)$ one can evaluate the expectation value as
\begin{equation}
  \int d\mu(T)\,f_{N^\prime}(T)=\int d\mu_{N^\prime}(T)\,f_{N^\prime}(T).
\end{equation}
It is convenient to parametrize the finite dimensional measures in terms of an action\footnote{The actions $S_{N^\prime}(T)$ are effective actions in the Wilsonian sense.} $S_{N^\prime}(T)$, such that for all $f_{N^\prime}(T)$
\begin{equation}\label{equ:FiniteDimensionalMeasureDefinition}
  \begin{array}{l}
  \int d\mu_{N^\prime}(T)\,f_{N^\prime}(T)\\[3mm] =\int \prod_{a=1}^{N^\prime}\prod_{b=1}^{N^\prime-a}\prod_{c=1}^{N^\prime-(a+b)}\,dT_{abc}\,f_{N^\prime}(T)\,\exp(-S_{N^\prime}(T)).
   \end{array}
\end{equation}
The actions $S_{N^\prime}(T)$ are all derived from the same measure $d\mu(T)$ on the infinite dimensional space.
Therefore, the family of actions $S_{N^\prime}(T)$ by construction satisfies the so-called {\it cylindrical consistency} relations the importance of which has also been emphasized for quantum gravity in \cite{Dittrich:2012jq}: 
\begin{equation}\label{equ:CylindricalConsiytencyRelations}
  \begin{array}{l}
  \quad\,\int \prod_{a=1}^{N^\prime}\prod_{b=1}^{{N^\prime}-a}\prod_{c=1}^{{N^\prime}-(a+b)}\,dT_{abc}\,f_{N^\prime}(T)\,\exp(-S_{N^\prime}(T))\\[5mm] =\int \prod_{a=1}^M\prod_{b=1}^{M-a}\prod_{c=1}^{M-(a+b)}\,dT_{abc}\,f_{N^\prime}(T)\,\exp(-S_M(T))
  \end{array}
\end{equation}
for all $M>N^\prime$ and all $f_{N^\prime}(T)$. In the more intuitive language applicable to the continuum, this is the statement that the low-energy dynamics arises by integrating out microscopic degrees of freedom, but that (sufficiently coarse) observables derived from the microscopic and the effective dynamics agree. In other words, cylindrical consistency is the statement that the family of actions all encode the same physics, just parameterized in different sets of degrees of freedom.

It turns out that the converse is also true: Given a family\footnote{Notice that we only require that there exists a measure for any finite $N^\prime$. It is a nontrivial mathematical fact that a cylindrically consistent family of finite dimensional measures defines a genuine measure on the infinite dimensional space that is given by all functions on the infinite dimensional space whose integrals are obtained as limits of Cauchy sequences of functions on finite dimensional spaces.} of measures on finite dimensional spaces $\{d\mu_{N^\prime}(T)\}_{N^\prime=0}^\infty$, 
encoded through the family of actions $\{S_{N^\prime}(T)\}_{N^\prime=0}^\infty$, these can be used to define a measure $d\mu(T)$ on the infinite dimensional space.  This requires that they satisfy the cylindrical consistency relation\footnote{The notion of cylindrical consistency that we consider here is a special case of the mathematical concept, which can be applied to any family of finite dimensional subspaces. In the general case one imposes a consistency relation analogous to Eq.~\eqref{equ:CylindricalConsiytencyRelations} when one finite dimensional subspace is contained in another ``larger" subspace.} Eq.~\eqref{equ:CylindricalConsiytencyRelations} for all $M>N^\prime$. That the limiting case with desired symmetries and locality properties exists is in general a nontrivial requirement, better known as the demand that a theory be asymptotically safe (or free).

It is useful to consider an expansion of the action $S(T)$ in terms of monomials
\begin{equation}
  S(T)=\sum_{i\in \mathcal I}g_i\,O_i(T),
\end{equation}
where $\mathcal I$ is an index set for a complete\footnote{Here we use the term complete in a loose sense, meaning that we can expand a arbitrary action functional with the correct symmetry properties in terms of the set of monomials. Note that this imposes a crucial assumption on theory space, which is that of quasilocal interactions in the continuum QFT case. For the tensor model case it consists in the assumption that all interactions can be expanded in positive powers of tensors. In the continuum QFT case this is critical for the Gaussian fixed point to only feature a finite number of relevant perturbations. We conjecture that a similar statement applies to the tensor model case.} set of tensor invariants that possess the symmetries that are imposed on the model. Cylindrical consistency is satisfied by the Gaussian integral with a trivial action. Physically, this is just the statement that a free theory remains free under coarse-graining.
An important nontrivial class of solutions to these cylindrical consistency relations are {\it self-similar} solutions, which take the form
\begin{equation}\label{equ:SelfSimilarSolution}
  S_{N^\prime}(T)=N^{\#}\sum_{i\in\mathcal I}g_i\,(N^\prime)^{d_i}(1+O(1/N^\prime))\,O_i(\Pi_{N^\prime}(T)),
\end{equation}
where $\Pi_{N^\prime}$ denotes the projection of the tensor onto a tensor  whose only non-vanishing components $T_{abc}$ satisfy $a+b+c\le N^\prime$.

Exact RG equations are essentially interpolation equations between effective actions $S_{N^\prime}(T)$ at different scales $N^\prime$. We thus intuitively expect that self-similar solutions can be found as fixed points of exact Renormalization Group equations and we plan to formalize this argument in detail in the future. 
An important step in the argument 
is the fact that the effective average action, which satisfies the FRGE, can be obtained as a (modified) Legendre transform of the Polchinski effective action \cite{Morris:1993qb}, which in turn satisfies the Polchinski equation. 
The Polchinski equation can be derived from the fact that Gaussian integrals, which are formally defined by $\int dG_A(T)e^{\langle J|T\rangle}:=\exp(\frac 1 2 \langle J |A^{-1}.J\rangle)$ satisfy\footnote{We will index Gaussian  measures $dG_A$ by the defining covariance matrix $A$. This may or may not involve a UV-cut-off $N^\prime$, depending on the covariance matrix used.} 
\begin{align}
 & \int dG_A(T)e^{\langle J|T\rangle}=\nonumber \int dG_C(T)dG_B(U)e^{\langle J|T+U\rangle}\\[3mm] & \textrm{ when }A^{-1}=B^{-1}+C^{-1}.
\end{align}
One can then use the covariance $B$ to interpolate between the effective actions by regularizing the propagator. In particular, the interpolation property for the Polchinski effective action can be shown to hold when $B$ turns into the kinetic term for $S_{N^\prime}(T)$ in one limit of the interpolation and into the kinetic term of $S_M(T)$ in the other limit of the interpolation. The Legendre transform to the effective average action relates the IR-suppression term of the FRG equation with the modification of the free kinetic term that is done in the setup of Polchinski's equation. This has important consequences, which will be important for the FRGE setup in the present paper:
\begin{enumerate}
\item The modified free kinetic term has to be compatible with the interpolation property and define a non-trivial Gaussian measure for all $N^\prime,M$.
\item Besides the interpolation property we need to ensure that the flow equations can be extended to $N^\prime,M$ arbitrarily large.
\item The kinetic term is held fixed in the derivation of the flow equation. This implies that we have to perform a scale- dependent redefinition of the tensor to absorb the flow of the normalization of the free kinetic term. This leads to the usual requirement that the bare couplings scale with the canonical plus anomalous dimension.
\end{enumerate}
These properties will be used to fix ambiguities that arise in the FRGE setup.

\subsection{Field Content and Symmetries}

In the present paper we are interested in pure tensor models that  
are defined as a 
path integral with a trivial kinetic operator, i.e., a quadratic term of the form
\be
S_{\mathrm{kin}} = \frac{1}{2}\sum_{a,b,c}T_{abc}T_{abc}\,.
\ee
To start with we consider large but finite tensors $T_{abc}$ and use the measure $[dT]$ which is obtained as the wedge product  of all $dT_{abc}$. In this section we will explicitly write all occurring sums, to indicate the finiteness of the tensors.  When reporting on practical calculations in the second part of the paper we will use the Einstein summation convention whenever it is unambiguous. To set up the theory space of our model, we start with the origin and consider perturbations about the free theory.
The Gaussian integral that we
aim at perturbing
is 
\begin{align}
  &\int [dT]\exp\left(-\frac{1}{2}\,\sum_{a,b,c}\,T_{abc}^2+\sum_{a,b,c}\,J^{abc}T_{abc}\right)\nonumber\\[3mm]&=\exp\left(\frac{1}{2}\,\sum_{a,b,c}\,(J^{abc})^2\right).
\end{align}
We impose that all interaction terms are invariant under the $O(N')^3$ action
\begin{equation}
 T_{abc}\mapsto O^{(1)}_{ad}O^{(2)}_{be}O^{(3)}_{cf}\,T_{def}\,.
\end{equation}
This symmetry is an important requirement for a well defined large-$N$ expansion, see \cite{Gurau:2011xp,Carrozza:2015adg,Gurau:2009tw,Gurau:2010ba,Gurau:2011xq,Gurau:2011aq,Bonzom:2012hw}. See, however \cite{Benedetti:2017qxl,Carrozza:2018ewt} for recent results indicating that this strong symmetry requirement might be dispensable.
The three ``copies" of the $O(N')$ transformations are distinct. Therefore, 
this symmetry implies that the first index can only be contracted with another first index, and similarly for the second and third index. For instance, contractions of the form $T_{abc}T_{bca}$ do not respect this symmetry (instead they only respect a reduced symmetry under which the indices at all positions transform simultaneously). Requiring the $O(N')^3$ symmetry is also sufficient to restrict the interactions to even powers in the tensors.\\
By analogy with local QFT, we restrict the basis in theory space to interactions with positive powers of tensors.
Thus we obtain a basis for the field monomials in terms of generalized traces, which are monomials of the tensor that can be written in the form
\begin{equation}
O^{\textrm{trace}}_i (T)=\overbrace{\delta^{a_1a_i}\delta^{b_1b_j}\delta^{c_1c_k}...}^{3n\,\delta\textrm{ s}}\overbrace{T_{a_1b_1c_1}...T_{a_{2n}b_{2n}c_{2n}}}^{2n\,\textrm{tensors}}.
\end{equation}
This means that we seek a self-similar solution of the form
\begin{equation}
S_{N'} (T)=N'^{\#}\sum_{i\in\mathcal I}g_i\,N'^{d_i}(1+O(1/N'))O^{\textrm{trace}}_i(T)
\end{equation}
as a fixed point of the effective average action $\Gamma_N (T)$ of the same form.

\subsection{IR-suppression Term} \label{IR-sup term}

To set up the functional RG equation, we introduce an IR suppression term into the generating functional. By IR we refer to small values of indices. In other words, fluctuations in the ``upper left" part of the tensors should be suppressed in the path integral, such that only the outermost ``layers" (what would be rows and columns in a matrix) are integrated out. In contrast to models on a background spacetime, there appears to be
an additional freedom in how this is implemented. To explain this, let us briefly review aspects of the IR suppression term in a background-dependent setting.

In the presence of a background spacetime, the canonical mass-dimensionality provides the scaling dimensions of all operators. The guiding principle for this is dimensional homogeneity: re-defining the unit of length (or equivalently mass) in all equations does not change their physical predictions, but simply re-expresses them in another set of units. Dimensional homogeneity implies that a quadratic IR-suppression term needs to carry the dimension of a mass-term.
Therefore, the ``natural" choice of cutoff term is
\be
\Delta_k S[\phi] = \frac{Z_{\phi}}{2}\, k^2\,\int_x \phi(x)R\left(\frac{\Delta}{k^2}\right)\phi(x)\,,\label{eq:cutoffbck}
\ee
with $\Delta = -\partial^2$ (or more generally $\Delta =-\bar{D}^2$ in the presence of a nontrivial background metric $\bar{g}_{\mu\nu}$). If one were to choose a different scaling with $k$, dimensional homogeneity required a compensation with another quantity with units of momentum. The only other quantity that exists, given that the IR-suppression term must be quadratic in the fields to ensure the one-loop structure of the flow equation, is the UV cutoff $\Lambda$, thus one would have
\begin{equation}\label{equ:StadnardIRsuppression}
\Delta_k S[\phi] = Z_{\phi}\frac{k^{2+r}}{2\,\Lambda^r}\int d^4x\,\phi(x)R\left(\frac{\Delta}{k^2}\right)\phi(x)\,.
 \end{equation}
We  
require
the beta functions to be independent of the UV cutoff $\Lambda$, so a change in $\Lambda$ does not affect them.
This independence of the RG equations 
from
$\Lambda$ allows us to always tacitly move $\Lambda$ in such a way that it is larger than $k$, so $\Lambda$ will never appear explicitly in the beta functions. Hence we can take the limit $k\to\infty$ without having to worry about the explicit appearance of $\Lambda$ in the beta functions.
Due to the requirement of independence from $\Lambda$, the unique choice for the scaling is Eq.~\eqref{eq:cutoffbck}. 

Let us now explore the background-independent tensor model setting.
The requirement of dimensional homogeneity, which results in dimensionless beta functions, can be translated into this setting by demanding that there is a large $N$ limit in which the beta functions are explicitly independent of $N$. Note that requiring autonomous beta functions can accordingly be used as a motivation to demand that there should be a well-defined $1/N$ expansion.
Yet some aspects of the situation are different in the background-free setting of tensor models. The IR-suppression term should also be quadratic in the tensors to ensure the one-loop structure of the flow equation. Hence, 
\be
\label{equ:PreviousIRsuppressionTerm}
  \Delta_NS[T]=\frac{Z_N}{2}\sum_{abc}T_{abc}^2\!\left(\!\frac{N}{a+b+c}-1\!\!\right)\theta\!\left(\!\frac{N}{a+b+c}-1\!\!\right), 
\ee
with a wave-function renormalization $Z_N$ is a ``natural" choice. Yet, the scaling with $N/(a+b+c)$ might not be the only viable option, as there is no mass-dimensionality of operators.  Thus, there is the more general choice
\be \label{equ:rIRsuppressionTerm}
\Delta_N^{(r)}S[T] = \frac{Z_N}{2}\sum_{abc}T_{abc}^2\!\left(\!\frac{N^r}{a+b+c}-1\!\!\right)\theta\!\left(\!\frac{N^r}{a+b+c}-1\!\!\right).
\ee
As there is no notion of mass dimensionality, all choices $r\geq 0$ appear permissible\footnote{The alternative choice $N^r \Delta_N S(T)$ results in the same beta functions as the choice $\Delta_N S(T)$, which is why we will not discuss it further. }. Ultimately, the strongest way of fixing $r$ is to consider the actual physics one is interested in exploring. In our case, we investigate tensor models with the aim of understanding quantum spacetime through dynamical triangulations. This allows us to fix the scaling dimension of one operator. In turn, this is sufficient to fix $r$, as we will see in subsec.~\ref{sec:ScalingFromGeometry}. In the following, we first discuss the assignment of scaling dimensions for operators in more detail before showing how the geometric interpretation of tensor models fixes $r$.

\subsection{Canonical Dimension}

A background spacetime automatically provides a canonical scaling dimension for all operators. This is a direct consequence of the fact that the standard RG flow
is literally a coarse-graining, i.e., a \emph{local} averaging process. Therefore the scaling of operators is related to their response under a rescaling of actual lengths/momenta. Accordingly, the canonical mass-dimensionality provides the canonical scaling. In background-independent models such as tensor models, there is no mass-dimensionality. Hence, this simple way of determining canonical scaling dimensions cannot be transferred from the background-dependent to the background-independent setting.
Yet, in local coarse-graining schemes, one can alternatively determine a consistent choice of scaling dimensions without prior knowledge of the canonical dimensionality by simply demanding that beta functions become autonomous, i.e., independent of the RG scale. This provides a unique assignment of scaling dimension which is in agreement with the mass-dimensionality, as expected.
This way of determining scaling dimensions is also available in tensor models. Specifically, the requirement that the beta functions admit a $1/N$-expansion, such that the leading $1/N^0$ terms are nontrivial, provides a way to fix the scaling dimension.
Demanding the $1/N$ expansion actually leads to upper bounds on the scaling dimensions. Choosing the scaling dimension for a coupling lower than the upper bound leads to a decoupling of that coupling from the system. Avoiding such artificial decoupling (which is an \emph{over}-suppression of couplings in the large $N$ limit) results in a unique choice of scaling dimension. The upper bounds depend on $r$ from Eq.~\eqref{equ:rIRsuppressionTerm}, which can be fixed by physical considerations, cf.~Sec.~\ref{sec:ScalingFromGeometry}. See \cite{Krajewski:2015clk,Krajewski:2016svb} for related discussions in the framework of the Polchinski equation.

In the separate paper \cite{EKLP2} we will show that one can also determine a scaling $(N^\prime)^{d_i}$ of all monomials $O_i (T)$ in a self-similar solution (\ref{equ:SelfSimilarSolution}) if one fixes the scaling of the quartic melon, i.e., the interaction
\begin{equation}
O_{\textrm{qm}}(T)=T_{abc}T_{dbc}T_{def}T_{aef},
\end{equation}
where summation over repeated indices is implied.

Before giving a brief outline of the argument let us observe that if we choose $d_{4m}=d_{\textrm{quartic melon}}$ too negative, then the suppression by the powers $(N^\prime)^{d_{4m}}$ in the continuum partition function
\begin{equation}
Z[J]=\lim_{N^\prime\to\infty}\int d \mu_{N { '}}(T)\exp\left(g\,{N^ \prime}^{d_{\textrm{4m}}}\,O_{\textrm{qm}}(T)+J\cdot T\right)
\end{equation}
will turn the partition function into a Gaussian partition function. On the other hand, a one-loop calculation shows that $d_{4m}=-2$ leads to a non-Gaussian partition function. Conversely, choosing $d_{4m}>-2$, the partition function will not converge in the limit $N^\prime\to\infty$. We therefore consider the critical value $d_{4m}=-2$ in the following. As we will see in the second part of this paper, this is exactly the value that is provided by inspecting the beta functions derived with the FRG setup.

 In \cite{EKLP2} we will show that a consistent scaling for all monomials of the rank three pure tensor model case is given by
\begin{equation}
  d(\gamma)=3-\frac{1}{2}\left(3p(\gamma)+f(\gamma)\right),
\label{scalingdim}
\end{equation}
where we used the fact that a tensor monomial $O_i(T)$ can be associated with a colored graph $\gamma$. This association is achieved by representing each tensor by a vertex and each contraction of an index between two tensors by a colored edge. The color of the edge is determined by the position of the index, so, e.g., first indices are assigned ``black", second indices ``blue" and third indices ``green." Then $2\,p(\gamma)$ is the number of vertices in $\gamma$ and $f(\gamma)$ is the number of faces of $\gamma$, where a face of $\gamma$ is a connected subgraph of $\gamma$ that contains only edges with two colors.

\subsection{Geometric interpretation of tensor models and geometric choice of scaling dimensions}\label{sec:ScalingFromGeometry}

Tensor models arise in various physical settings, in particular as a tool to attempt a definition of  
the quantum-gravity partition function. The particular physical model at hand dictates the scaling of the bare action with the tensor size $N$. This scaling can be taken as an initial condition to find the canonical scaling. For simplicity we present the argument for colored tensor models \cite{Gurau:2011xp}, which determine, after integration over all but one color, the scaling of the cyclic melon.

The connection with the gravity partition-function is made by considering the continuum limit of discretizations of the Einstein-Hilbert action $S=\frac{1}{16\,\pi\,G}\int d^Dx\sqrt{|g|}(R[g]-2\Lambda)$ in $D$ dimensions. 
Herein $G$ and $\Lambda$ denote the (dimensionful) Newton coupling and cosmological constant, respectively. The probably simplest approach consists of considering statistical models of triangulations $\Delta$ in terms of equilateral simplices of size $a$. The continuum limit is taken by sending the lattice size $a\to 0$ while holding the expectation value of the volume fixed. 
The $D$- simplices of the triangulation $\Delta$ are assumed to be flat and curvature is associated with $D-2$ dimensional hinges in $\Delta$. The corresponding form of the Einstein-Hilbert action 
reads
\begin{equation}\label{equ:ReggeAction}
    S_{R}[\Delta]=\kappa_D\,N_D(\Delta)-\kappa_{D-2}\,N_{D-2}(\Delta),
\end{equation}
where $N_D(\Delta)$ denotes the number of $D$-simplices in $\Delta$ and $N_{D-2}(\Delta)$ the number of $(D-2)$- simplices (hinges) in $\Delta$. One finds from Regge calculus that $\kappa_{D-2}\propto 1/G$ and $\kappa_D$ is a linear combination of $\Lambda/G$ and $1/G$.
The grand-canonical\footnote{One usually refers to this version of the gravity partition function as grand-canonical because it is, analogous to the grand-canonical partition function in statistical mechanics, formulated for a fluctuating number of building blocks.} gravity partition function is then
\begin{equation}
    Z_{grav}=\sum_\Delta\frac{e^{-S_R[\Delta]}}{S[\Delta]},
\end{equation}
where $S[\Delta]$ denotes the size of the (discrete) symmetry group of the triangulation $\Delta$. The volume $V[\Delta]$ of a triangulation $\Delta$ is given by $V[\Delta]=a^D\,V_D\,N_D$, where $V_D$ is the volume of a unit equilateral $D$-simplex. Thus, one obtains the expectation value of the volume by 
\begin{equation}
    \langle V \rangle =-a^D\,V_D\,\frac{\partial}{\partial\,\kappa_D}\ln(Z_{grav}).
\end{equation}
This 
we want to hold at a fixed value $V_o$ when taking the limit $a\to 0$. Consequently, we take the continuum limit in such a way  that
\begin{equation}\label{equ:LimitAto0}
 \textrm{const.}=\lim_{a\to 0}\,\left[a^D\frac{\partial}{\partial \,\kappa_D}\ln(Z_{grav}(\kappa_D(a),\kappa_{D-2}(a)))\right],
\end{equation}
where we anticipate the RG-scaling of the combinatorial couplings $\kappa_D$ and $\kappa_{D-2}$ (i.e., their dependence on the lattice spacing $a$) and the constant is associated with $V_o/V_D$. 
Thus the logarithmic derivative of the gravity partition function w.r.t. the combinatorial coupling $\kappa_D$ has to scale with $a^{-D}$ in the limit $a\to 0$  
in order for
the continuum limit to exist. This is automatically satisfied if
$\kappa_D(a)$ assumes the dimensional scaling of the cosmological constant $\Lambda/G$ with the lattice size $a$. However, if one considers a purely combinatorial approach to quantum gravity where no lattice size is introduced before hand, then one has only the combinatorial couplings at ones disposal. 
Hence, to be able to associate the continuum limit of  
the quantum gravity partition function with one of these models, one has to find a point of the partition function where $\frac{1}{Z_{grav}}\frac{\partial\,Z_{grav}}{\partial\,\kappa_D}$  diverges, to be able to maintain $\langle V\rangle$ fixed while taking $a\to 0$. This means that we are looking for a critical point of the partition function, where we can tune $\kappa_D$ in such a way that we can take the fixed-volume limit as implied in Eq.~\eqref{equ:LimitAto0}.

We will now outline the argument explaining why certain tensor-model partition functions possess an interpretation as combinatorial discrete gravity partition functions.  Hence the search for a continuum limit of quantum gravity can be  
tackled
by studying critical points of the tensor-model partition functions. We will start with colored tensor models where the connection with dynamical triangulation is direct.  
Next, we will
apply the observations made in these models to consider uncolored tensor models that are formally obtained by integrating out all but the last colored tensor.

The partition function of the colored rank $D=3$ tensor model is based on the bare action: 
\bea
   S(T,\bar T)&=& \sum^3_{i=0}T^i_{abc}\bar{T}^i_{abc}\nonumber\\
   &{}&+N^{-3/2}\left(\lambda\, T^0_{abc}T^1_{ade}T^2_{fbe}T^3_{fdc}+c.c.\right)\,,
\eea
where $\bar T$ stands for the complex conjugate of $T$. The Feynman diagrams of this model are built from vertices and colored edges. It can be shown \cite{Gurau:2010nd} that each Feynman diagram is dual to a finite simplicial pseudo-manifold (i.e., a triangulation): vertices are dual to tetrahedra, edges dual to triangles and faces (closed two-colored subgraphs) are dual to lines. The scaling of the interaction with $N^{-3/2}$ is important for the existence of a $1/N$ expansion for the $n$-point functions \cite{Gurau:2010ba}. With this scaling one obtains the amplitude of a Feynman graph $\Gamma$ as
\begin{equation}\label{equ:AmplitudeGamma}
  A(\Gamma)=N^{N_{1}-3/2N_3}\,(\lambda\bar\lambda)^{N_3/2},
\end{equation}
where $N_3$ denotes the number of vertices and $N_{1}$ the number of faces in the triangulation $\Delta(\Gamma)$ that is dual to $\Gamma$. For the physical interpretation of tensor models as encoding dynamical triangulations, one should compare with the Regge amplitude 
\begin{equation}\label{equ:AmplitudeRegge}
A(\Delta)=e^{-S_R[\Delta]}=\exp(\kappa_{1}N_{1}-\kappa_3N_3)
\end{equation}
 of a triangulation by equilateral building blocks. Comparing equations (\ref{equ:AmplitudeGamma}) and (\ref{equ:AmplitudeRegge}) leads to the identification of the combinatorial couplings as
\bea
  \kappa_{2}&=&\ln(N),\nonumber\\
  \kappa_3&=&\frac 3 2\ln(N)-\frac 1 2 \ln(\lambda\bar\lambda).
\eea
The dualization of Feynman graphs $\Gamma$ to triangulations $\Delta(\Gamma)$ allows us to identify the partition function $Z$ of the colored tensor model with the sum over all (possibly disconnected) triangulations $\Delta^\prime$ weighted by a Boltzmann factor associated with the Regge action. We can thus identify the free energy $F=\ln(Z)$ of the tensor model with the grand canonical partition function $Z_{grav}$ of dynamical triangulations
\begin{equation}
  F=\sum_{\Delta}\frac{1}{S(\Delta)}\exp(\kappa_{1}N_{1}-\kappa_3N_3)=Z_{grav},
\end{equation}
where $S(\Delta)$ denotes the size of the symmetry group of the triangulation $\Delta$.

It is important to remember that the dualization of the tensor-model free energy $F$ to the quantum-gravity partition function $Z_{grav}$ requires the scaling $N^{-3/2}$ for the tetrahedral interaction\footnote{Notice that the complex colored model possesses a tetrahedral interaction, which is not present in the complex uncolored model.}. After integrating out all colors except color 0 in the model, one obtains a complex version of the uncolored tensor model \cite{Bonzom:2012hw} of the kind that we are investigating in the present paper. The scaling of the coupling of the cyclic 4-melon is determined by the scaling of the original colored model. The result of this translation is a scaling of the coupling with $N^{-2}$. This is the initial condition for finding the canonical dimensions that ensures the interpretation of the RG fixed points in terms of continuum limits of dynamical triangulations. In turn, it is actually the scaling dimension that the flow equation provides as the upper bound on the scaling of the coupling such that the beta functions admit a $1/N$ expansion. To summarize, even without prior knowledge on scaling dimensions, the FRG provides upper bounds on these from the requirement of a well-defined $1/N$ expansion once $r=1$ is set in the regulator in Eq.~\eqref{equ:rIRsuppressionTerm}.  
In summary,
the scaling dimension of the quartic coupling can  
be fixed by physical considerations, specifically the interpretation of tensor models as dual to triangulations. A nontrivial continuum limit in gravity then fixes the scaling dimension, in agreement with the upper bound inferred by inspection of the beta functions for that choice of $r$ in which the setup is closest to the analogous setup for local coarse-graining, i.e., $r=1$.

\section{The model}\label{sec:model}
In the remainder of this work we consider a real rank-3 tensor model. It  is characterized by $O (N^\prime)\otimes O (N^\prime)\otimes O (N^\prime)$ 
symmetry,  such that the action is invariant under
\be\label{model1}
T_{a_1 a_2 a_3}\,\,\rightarrow\,\, T^{\prime}_{a_1 a_2 a_3} = O^{(1)}_{a_1 b_1} O^{(2)}_{a_2 b_2} O^{(3)}_{a_3 b_3} T_{b_1 b_2 b_3}\,.
\ee
Herein,
$O^{(i)}\,,\, i=1,2,3$ are independent orthogonal matrices. There is no symmetry under interchanges of the indices, restricting the way one can contract tensors. In particular, the first index of a given tensor must be contracted with the first index of 
another tensor and so on. This can be represented by assigning ``colors" to each index position to emphasize the distinct role they play. This structure implies that besides the interactions which are present 
in the combinatorial sense 
in the rank-3 complex tensor model investigated in  \cite{Eichhorn:2017xhy}, new interactions are allowed by the orthogonal symmetry.
In particular, in the complex model, a tensor $\mathbb{T}$ and its complex conjugate $\bar{\mathbb{T}}$ are introduced and invariance under unitary transformations implies that tensors $\mathbb{T}$ must be contracted with $\bar{\mathbb{T}}$,  i.e., interaction terms must contain the same number of $\mathbb{T}$'s and $\bar{\mathbb{T}}$'s. In the real case, there is just one type of tensor, namely $T$, allowing for new combinatorial structures.

\subsection{Truncation}
Our ansatz for the effective average action $\Gamma_N$ includes all $O (N^\prime)\otimes O (N^\prime)\otimes O (N^\prime)$ invariant interactions\footnote{Note that $N'$ is the UV cutoff and distinct from the IR cutoff $N$, on which the effective average action depends.} up to sixth order in $T$. 
The principle underlying our truncation is the assumption that canonical ordering provides a useful guide to set up robust truncations.
Increasing numbers of tensors in an interaction result in lower canonical dimension of the corresponding coupling 
\footnote{Combinatorically distinct interactions at a given number of tensors differ in their canonical dimensions. Therefore, the most relevant interaction at eighth order in $T$'s is in fact less irrelevant than particular interactions at $T^6$. For practical purposes, we nevertheless do not include $\mathcal{O}(T^8)$ interactions.}.  Our truncation of the effective average action is 
accordingly chosen to be
\begin{equation}
\Gamma_N = \Gamma_{N,\,2} + \Gamma_{N,\,4} + \Gamma_{N,\,6}\,,
\label{model2}
\end{equation}
with all parts represented diagrammatically in Fig.~\ref{trunc} and 
\bea
\Gamma_{N,\,2}& =& \frac{Z_N}{2}T_{a_1 a_2 a_3}T_{a_1 a_2 a_3}\,,
\label{model3}\\
\Gamma_{N,\,4}&=& \bar{g}^{2,1}_{4,1}\,T_{a_1 a_2 a_3}T_{b_1 a_2 a_3}T_{b_1 b_2 b_3} T_{a_1 b_2 b_3}\nonumber\\
&+&\bar{g}^{2,2}_{4,1}\,T_{a_1 a_2 a_3}T_{a_1 b_2 a_3}T_{b_1 b_2 b_3} T_{b_1 a_2 b_3}\nonumber\\
&+& \bar{g}^{2,3}_{4,1}\,T_{a_1 a_2 a_3}T_{a_1 a_2 b_3}T_{b_1 b_2 b_3} T_{b_1 b_2 a_3}\nonumber\\
&+& \bar{g}^0_{4,1}\,T_{a_1 a_2 a_3}T_{b_1 a_2 b_3}T_{b_1 b_2 a_3}T_{a_1 b_2 b_3}\nonumber\\
&+&\bar{g}^2_{4,2}\,T_{a_1 a_2 a_3}T_{a_1 a_2 a_3}T_{b_1 b_2 b_3}T_{b_1 b_2 b_3}\,,
\label{model4}
\eea
and
\begin{eqnarray}
\Gamma_{N,\,6} &=&  \bar{g}^{3,1}_{6,1}\,T_{a_1 a_2 a_3}T_{b_1 a_2 a_3}T_{b_1 b_2 b_3}T_{c_1 b_2 b_3}T_{c_1 c_2 c_3}T_{a_1 c_2 c_3} \nonumber\\
&+& \bar{g}^{3,2}_{6,1}\,T_{a_1 a_2 a_3}T_{a_1 b_2 a_3}T_{b_1 b_2 b_3}T_{b_1 c_2 b_3}T_{c_1 c_2 c_3}T_{c_1 a_2 c_3}\nonumber\\
&+&\bar{g}^{3,3}_{6,1}\,T_{a_1 a_2 a_3}T_{a_1 a_2 b_3}T_{b_1 b_2 b_3}T_{b_1 b_2 c_3}T_{c_1 c_2 c_3}T_{c_1 c_2 a_3}\nonumber\\
&+&\bar{g}^{2,1}_{6,1}\,T_{a_1 a_2 a_3} T_{a_1 b_2 a_3} T_{b_1 b_2 b_3} T_{c_1 c_2 b_3} T_{c_1 c_2 c_3} T_{b_1 a_2 c_3}\nonumber\\
&+& \bar{g}^{2,2}_{6,1}\, T_{a_1 a_2 a_3}T_{b_1 a_2 a_3}T_{b_1 b_2 b_3}T_{c_1 c_2 b_3}T_{c_1 c_2 c_3}T_{a_1 b_2 c_3}\nonumber\\
&+&\bar{g}^{2,3}_{6,1}\, T_{a_1 a_2 a_3} T_{b_1 a_2 a_3} T_{b_1 b_2 b_3} T_{c_1 b_2 c_3} T_{c_1 c_2 c_3} T_{a_1 c_2 b_3}\nonumber\\
&+& \bar{g}^{1,1}_{6,1}\, T_{a_1 a_2 a_3} T_{a_1 b_2 b_3} T_{b_1 b_2 a_3} T_{b_1 c_2 c_3} T_{c_1 c_2 c_3} T_{c_1 a_2 b_3}
\nonumber\\
&+& \bar{g}^{1,2}_{6,1}\, T_{a_1 a_2 a_3} T_{b_1 a_2 b_3} T_{b_1 b_2 a_3} T_{c_1 b_2 c_3} T_{c_1 c_2 c_3} T_{a_1 c_2 b_3}\nonumber\\
&+&\bar{g}^{1,3}_{6,1}\,T_{a_1 a_2 a_3} T_{b_1 b_2 a_3} T_{a_1 b_2 b_3} T_{c_1 c_2 b_3} T_{c_1 c_2 c_3} T_{b_1 a_2 c_3}
\nonumber\\
&+& \bar{g}^{0,np}_{6,1}\, T_{a_1 a_2 a_3} T_{b_1 a_2 b_3} T_{b_1 c_2 c_3} T_{c_1 c_2 a_3} T_{c_1 b_2 b_3} T_{a_1 b_2 c_3}\nonumber\\
&+&\bar{g}^{0,p}_{6,1}\,T_{a_1 a_2 a_3}T_{b_1 a_2 b_3} T_{b_1 b_2 c_3}T_{c_1 b_2 b_3}T_{c_1 c_2 a_3}T_{a_1 c_2 c_3}
\nonumber\\
&+& \bar{g}^{3,1}_{6,2}\,T_{a_1 a_2 a_3} T_{b_1 a_2 a_3} T_{b_1 b_2 b_3} T_{a_1 b_2 b_3} T_{c_1 c_2 c_3} T_{c_1 c_2 c_3} 
\nonumber\\\
&+& \bar{g}^{3,2}_{6,2}\,T_{a_1 a_2 a_3} T_{a_1 b_2 a_3} T_{b_1 b_2 b_3} T_{b_1 a_2 b_3} T_{c_1 c_2 c_3} T_{c_1 c_2 c_3}
\nonumber\\
&+& \bar{g}^{3,3}_{6,2}\,T_{a_1 a_2 a_3} T_{a_1 a_2 b_3} T_{b_1 b_2 b_3} T_{b_1 b_2 a_3} T_{c_1 c_2 c_3} T_{c_1 c_2 c_3} \nonumber\\
&+& \bar{g}^{1}_{6,2}\,T_{a_1 a_2 a_3} T_{b_1 a_2 b_3} T_{b_1 b_2 a_3} T_{a_1 b_2 b_3} T_{c_1 c_2 c_3} T_{c_1 c_2 c_3}
\nonumber\\
&+& \bar{g}^{3}_{6,3}\,T_{a_1 a_2 a_3} T_{a_1 a_2 a_3} T_{b_1 b_2 b_3} T_{b_1 b_2 b_3} T_{c_1 c_2 c_3} T_{c_1 c_2 c_3}\,.\nonumber\\
\label{model5}
\end{eqnarray}

\begin{widetext}
\begin{center}
\begin{figure}
\includegraphics[scale=.3]{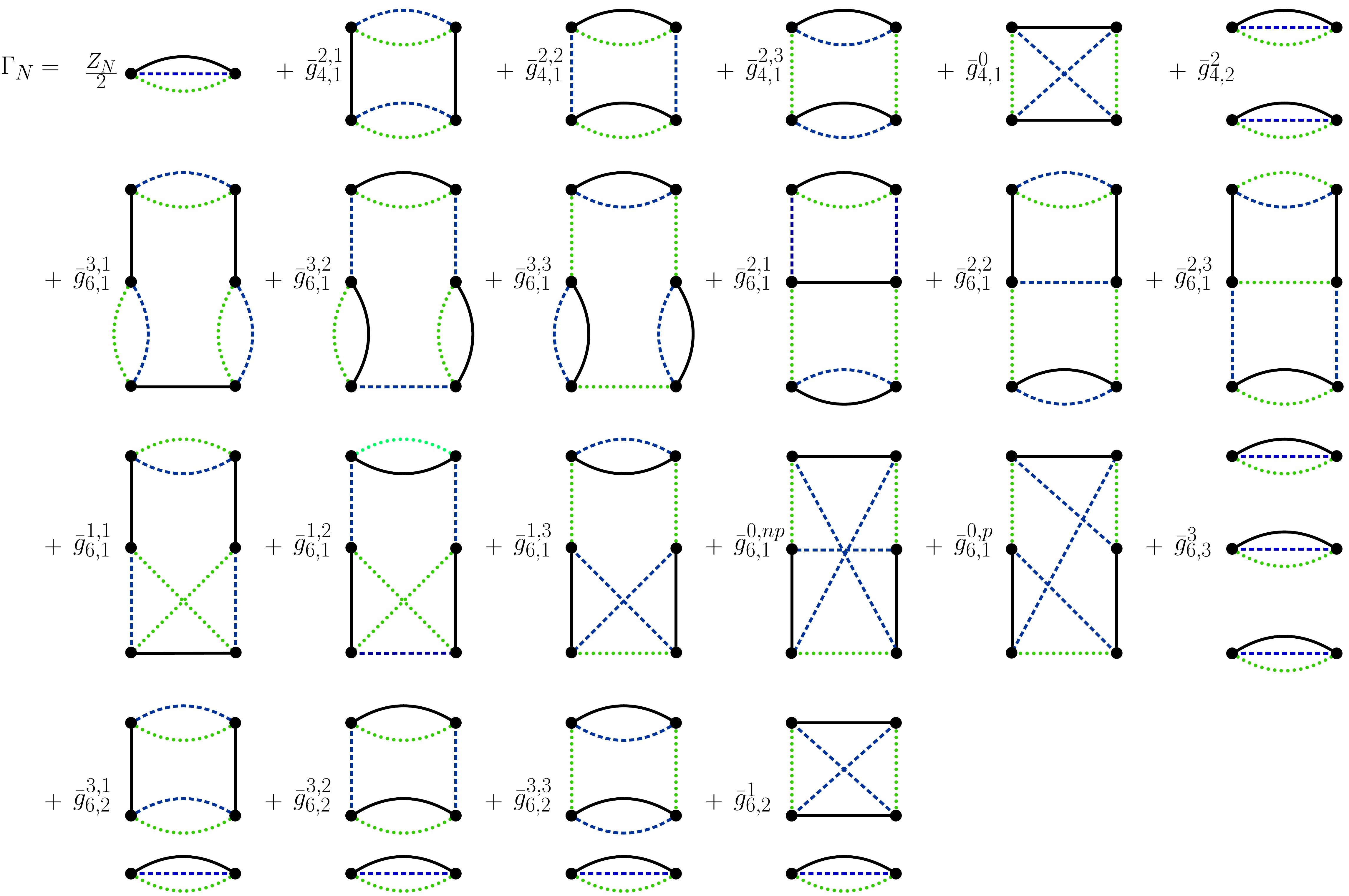}
\caption{Our truncation includes interactions up to $T^6$ invariant under $O (N^\prime)\otimes O (N^\prime)\otimes O (N^\prime)$ transformations. Black dots denote the tensors and each line that links different dots represents an index contraction. Different colors and linestyles distinguish different index positions.}
\label{trunc}
\end{figure}
\end{center} 
\end{widetext}
Here, $Z_N$ represents the wave-function renormalization and we employ the same notation for the couplings $\bar{g}^{i,j,\ldots}_{k,l,\ldots}$ as  
introduced in \cite{Eichhorn:2017xhy}: Of the two lower indices, the first counts the number of tensors and the second the number of connected components in the diagrammatic representation. 

For the rank-2 case, the second index accordingly counts the number of traces included in an interaction. Of the upper indices, the first counts the number of sub-melons. The second index is used to indicate a distinct color, whenever it takes the value 1, 2 or 3.
Additionally, we introduce a new structure in the notation: at sixth order in $T$, there are two types of interactions not containing
any sub-melons, as indicated by the first upper index taking the value 0. They are denoted by $g^{0,np}_{6,1}$ and $g^{0,p}_{6,1}$. The super-indices $p$ or $np$ indicate whether the graph corresponding to the interaction vertex is planar or not. In the complex model discussed in \cite{Eichhorn:2017xhy} unitary invariance implies that at sixth order in $T$ diagrams without sub-melons must be non-planar.
Let us comment on the presence of the multitrace interactions \footnote{This name is due to the matrix models analogue: in such models, interactions of the type $(\mathrm{Tr}(\phi^n))^m$, namely, product of traces, are also allowed by symmetry, where $\phi_{ab}$ are the random matrices.} $T_{a_1 a_2 a_3}T_{a_1 a_2 a_3}T_{b_1 b_2 b_3}T_{b_1 b_2 b_3}$, $T_{a_1 a_2 a_3} T_{a_1 a_2 a_3} T_{b_1 b_2 b_3} T_{b_1 b_2 b_3} T_{c_1 c_2 c_3} T_{c_1 c_2 c_3}$ etc. in the action. Combinatorically, these are disconnected interactions.  
One might at a first glance not consider 
their inclusion in a truncation necessary when considering the analogous case of background-dependent QFTs.
In standard local QFTs on a  
background
spacetime, one finds that the effective action does not contain the analogous disconnected terms $(\int d^4x_1 L_1(x_1))...(\int d^4x_n L_n(x_n))$. This is due to the fact that  
such
non-local terms are suppressed by inverse powers of the spacetime volume and hence vanish in the infinite volume limit. The RG reflects this property by not generating such terms from a local truncation \footnote{Non-localities in the effective action arise as resummations of quasi-local terms, see, e.g., \cite{Codello:2010mj}. If nonlocal terms are explicitly inserted in theory space they prevent the existence of a predictive continuum limit, as already the Gaussian fixed point features an infinite number of relevant perturbations.}.
The absence of spacetime locality in tensor models changes the argument and the analogue of non-local operators appears from the onset. In particular, the  RG flow generates tensor monomials whose indices are contracted according to a disconnected graph. It is important to point out that such disconnected graphs do not represent disconnected spacetimes, but are obtained as disconnected boundaries of a connected spacetime, such as, e.g., the spacetime between two compact spatial Cauchy surfaces.

To extract beta functions from the FRG
we have to specify a regulator function $R_N$. Following the discussion in subsection \ref{IR-sup term} and extending the setup for matrix models in \cite{Eichhorn:2013isa} and tensor models \cite{Eichhorn:2017xhy}, we introduce the regulator as
\begin{eqnarray}
R_N (\left\{a_i\right\},\left\{b_i\right\}) &=& Z_N \,\delta_{a_1 b_1}\delta_{a_2 b_2} \delta_{a_3 b_3} \left(\frac{N^r}{a_1+a_2+a_3}-1\right) \nonumber\\
&\times&\theta \left(\frac{N^r}{a_1+a_2+a_3}-1\right)\,,
\label{model6}
\end{eqnarray} 
A key ingredient in the flow equation is the scale-derivative of the regulator
\begin{eqnarray}
\partial_t R_N (\left\{a_i\right\},\left\{b_i\right\}) &=& \delta_{a_1 b_1}\delta_{a_2 b_2} \delta_{a_3 b_3} \left[N^r\frac{\partial_t Z_N+Z_N\, r}{a_1+a_2+a_3}\right.\nonumber\\
&-&\left.\partial_t Z_N\right]\theta \left(\frac{N^r}{a_1+a_2+a_3}-1\right)\,,
\label{model7}
\end{eqnarray}
with $\partial_t \equiv N\partial_N$. For $r>0$, the regulator \eqref{model6} plays the role of a ``momentum-dependent" mass term and suppresses the integration over modes for which $a_1+a_2+a_3 < N$. Accordingly, only the outermost 
``layers"
of the tensor $T_{abc}$ are integrated out. Thus, the orthogonal invariance $O(N^\prime)\otimes O(N^\prime) \otimes O(N^\prime)$ is necessarily broken by the introduction of the regulator, irrespective of the choice of shape function.

\subsection{Assignment of canonical dimension}

As discussed above, background independence enjoyed by tensor models makes the task of assigning canonical dimensions to the couplings  
non-trivial and non-unique unless additional conditions are imposed. In conventional quantum field theory the canonical \emph{scaling} dimensionality of couplings or operators under an RG step follows from the canonical \emph{mass} dimensionality, as the RG step is a local coarse graining step. This is no longer possible in background- and hence scale-free theories such as tensor models, because no mass dimensionality exists - all elements of the action are dimensionless from the outset. 

In order to find a self-consistent way to fix the canonical scaling dimension under changes in the ``scale" $N$ for the couplings in the rank-3 real tensor model, we use the strategy employed in previous works \cite{Eichhorn:2017xhy,Benedetti:2014qsa,BenGeloun:2018ekd}: We compute the beta functions for couplings $\bar{g}^{i,j,\ldots}_{k,l,\ldots}$ which we refer to as ``dimensionful" couplings from now on. This does not refer to an actual mass dimensionality, but simply borrows the standard terminology and means that they are not yet rescaled with an appropriate power of $N$.
The corresponding beta functions form a non-autonomous system, i.e., feature an explicit dependence on $N$, cf.~App.~\ref{explicit forms of beta functions}. Next, we impose that the system features a $1/N$ expansion with non-trivial terms at leading order $1/N^0$. This is achieved by rescaling all couplings by a suitable power of $N$ which can be read off the beta functions for the dimensionful couplings. This rescaling defines their canonical dimension. As we are interested in the large $N$ limit, we only keep the $1/N^0$ terms after the rescaling. This results in an autonomous system of beta functions.
 
Actually, requiring that the system allows a $1/N$ expansion is not enough to uniquely fix the scaling dimensions: Typically, demanding the existence of a well-defined and non-trivial large-$N$ limit only results in upper bounds for the scaling dimensions of couplings at least if one works in truncations. It is still consistent with the $1/N$ expansion to only saturate the upper bound for some couplings but not for all, see, e.g., \cite{Eichhorn:2017xhy,BenGeloun:2018ekd}. A choice of canonical dimension below the upper bound leads to a decoupling of the corresponding interaction in the system of beta functions. Accordingly, even the anomalous dimension  $\eta=-\partial_t Z_N/Z_N$ becomes trivial if \emph{none} of the canonical dimensions saturate the upper bound.

It should be emphasized that the large-$N$ limit plays a key role in the assignment of dimension. In particular, it is typically impossible to assign a consistent canonical scaling to all couplings in such a way that the system of beta functions becomes autonomous at finite $N$. This is a consequence of each term in the flow equation coming with several contributions that are subleading at large-$N$. Accordingly, if one analyzes regimes other than the large-$N$ one (e.g., $N \approx 1$), the assignment of  canonical dimensionality which leads to a non-trivial beta function system is different from the one obtained at large $N$, see, e.g., \cite{Benedetti:2014qsa}.\\

Let us emphasize a key difference to standard QFTs, where couplings associated to local interaction terms with the same number of fields and derivatives have the same canonical dimension irrespective of their internal index structure. In the present framework, 
the non-trivial combinatorial structure of the interactions entails different upper bounds on the scalings with $N$ even for interactions with the same number of tensors $T$. 
It is actually not possible to choose canonical dimensions below the upper bounds in such a way that interactions with the same number of tensors feature the same dimension. The problem with such a strategy is twofold: the canonical dimension in the melonic sector is fixed, instead of just satisfying an upper bound as we demand that the anomalous dimension should be nontrivial. Therefore, all other interactions would have to share the same canonical dimension. Yet, for several combinatorial structures, the upper bound on the canonical dimension lies below that for the melonic interactions.  Thus, it is impossible to set all canonical dimensions equal to each other at a given order in tensors.

In order to use the procedure described above in a concrete way, we choose a normalization condition such that the wave-function renormalization  $Z_N$ is dimensionless. The next step is the calculation of the anomalous dimension $\eta$ from the flow equation. Due to the one-loop structure of the flow equation, the interaction terms that contribute to the anomalous dimension are the quartic ones. Hence by demanding that $\eta$ features a well-defined large-$N$ limit, we can set bounds for the canonical dimensions of the quartic couplings. Then, the procedure extends to the quartic and sixth order couplings by inspecting their beta functions as explained in \cite{Eichhorn:2017xhy}. As discussed before, the large-$N$ expandability criterion does not fix  the canonical dimension uniquely but only sets an upper bound for almost all couplings. The  smaller, compared to the upper bound, the canonical dimension is chosen, the more suppressed the corresponding coupling becomes in the system of beta functions. For this reason, in this work we take the canonical dimensions to be the upper bounds. For the regulator \eqref{model6}  the upper bounds are:
\begin{align}
\left[\bar{g}^{2,i}_{4,1}\right] &= -2r \quad&\left[\bar{g}^{0}_{4,1}\right] &= -3r/2 \quad&&\left[\bar{g}^{2}_{4,2}\right] &= -3r\nonumber\\
\left[\bar{g}^{1,i}_{6,1}\right] &=-7r/2 \quad&\left[\bar{g}^{2,i}_{6,1}\right] & = -4r \quad&&\left[\bar{g}^{3,i}_{6,1}\right] &=-4r\nonumber\\
\left[\bar{g}^{0,np}_{6,1}\right] &=-3r\quad &\left[\bar{g}^{0,p}_{6,1}\right]& = -3r \quad &&\left[\bar{g}^{3,i}_{6,2}\right]& = -5r \nonumber\\
\left[\bar{g}^{1}_{6,2}\right] &= -9r/2\quad &\left[\bar{g}^{3}_{6,3}\right]& = -6r \,.\label{eq:lowerbounds}
\end{align}
 Lower bounds also exist for some couplings in our truncation and are expected to arise for further couplings beyond our truncation.  An important case is that 
 of $[\bar{g}^{2}_{4,2}]$ and $[\bar{g}^{3}_{6,3}]$, where the lower bounds coincide with the upper bounds. Therefore, for such couplings, the canonical dimensions are completely determined once the upper bound for the quartic cyclic melonic coupling, $g^{2,i}_{4,1}$, is saturated. In practice, this works as follows: Demanding that the anomalous dimension (\ref{fulletar}) has a non-trivial large-$N$ expansion implies $[\bar{g}^{2}_{4,2}]\le -3r$. On the other hand, the requirement that the beta function for $g^{2}_{4,2}$, see Eq.~\eqref{fullbetag242r}, has a well defined large-$N$ expansion leads to 
\be
2[\bar{g}^{2,i}_{4,1}]-[\bar{g}^{2}_{4,2}]+r \leq 0\,.
\label{upperquarticdisc}
\ee
Once the canonical dimension $[\bar{g}^{2,i}_{4,1}]$ is chosen to saturate the upper bound given in \eqref{eq:lowerbounds}, one obtains $[\bar{g}^{2}_{4,2}]\ge -3r$ implying $[\bar{g}^{2}_{4,2}]=-3r$. Similarly, the $1/N$-expandability of the beta function for $g^{2}_{4,2}$  
provides
a lower bound for $[\bar{g}^{3}_{6,3}]$, i.e., 
\be
[\bar{g}^{3}_{6,3}]-[\bar{g}^{2}_{4,2}]+3r\leq 0\,.
\label{lowebounddiscsix}
\ee
The beta function for $g^{3}_{6,3}$, see Eq.~\eqref{fullbetag363r}, has a well-defined and non-trivial large-$N$ limit if
\be
[\bar{g}^{2,i}_{6,1}]+[\bar{g}^{2}_{4,2}]-[\bar{g}^{3}_{6,3}]\leq 0\,.
\label{upbounddiscsix}
\ee
Plugging the value for $[\bar{g}^{2}_{4,2}]$ and $[\bar{g}^{2,i}_{6,1}]$ (which can be completely fixed as well) in \eqref{lowebounddiscsix} and \eqref{upbounddiscsix} gives $[\bar{g}^{3}_{6,3}]=-6r$. This shows that it is not possible to assign canonical dimensions for such multitrace interactions lower than the upper bounds as a strategy to decouple them from the other beta functions.

\subsection{Projection rule}
To extract the beta functions, we must project the flow onto the corresponding directions in theory space. To that end, we need to decompose the flow of $\Gamma_N$ into the basis monomials that span the theory space.  In our case, we only project onto the subspace spanned by the 21 couplings that define our truncation. 
The $O(N')\otimes O(N')\otimes O(N')$ symmetry-breaking induced by the regulator is an added complication. It leads to the generation of additional interactions which are not invariant under the $O(N')\otimes O(N')\otimes O(N')$ symmetry. This is manifest in their explicit index dependence. An inspection of the right-hand-side of the flow equation immediately reveals the presence of such terms. Our task therefore consists in isolating distinct combinatorial structures at each order in the tensors, but also of isolating symmetry-invariants and symmetry-breaking terms.

\begin{figure}[!t]
	\includegraphics[width=\linewidth,clip=true,trim=0cm 6cm 0cm 0.5cm]{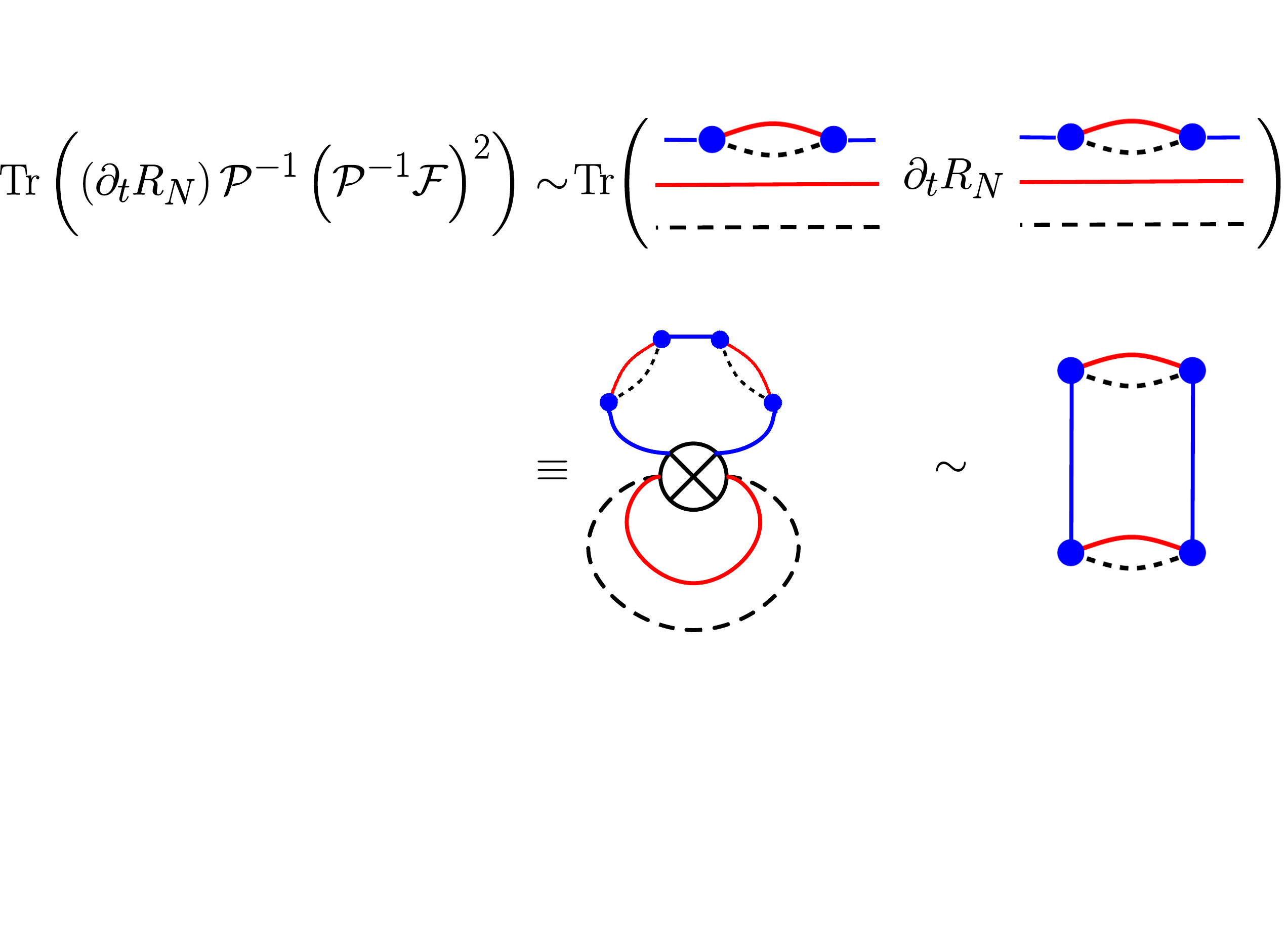}
	\caption{\label{fig:PFexpansion}Depicted above is the contribution of $\left(g^{2,1}_{4,1}\right)^2$ to the running of $g^{2,1}_{4,1}$ arising at second-order in the $\mathcal{P}^{-1}\mathcal{F}$ expansion. The diagrammatic terms in the first row originate from computing the field-dependent two-point function, i.e., $\mathcal{F}$. In the lower diagram on the left, we have replaced the scale-derivative of the regulator by a cross and contracted all lines. }
\end{figure}

A possible strategy for the calculations with the flow equation is to use the $\mathcal{P}^{-1}\mathcal{F}$ expansion, i.e., the flow equation is expressed as
\bea
\partial_t \Gamma_N& =& \frac{1}{2}\mathrm{Tr}(\partial_t R_N)\mathcal{P}^{-1}\nonumber\\
&{}&+\frac{1}{2}\sum^{\infty}_{n=1}(-1)^n {\rm Tr}\left[(\partial_t R_N)\mathcal{P}^{-1} (\mathcal{P}^{-1}\mathcal{F})^n\right]\,,
\label{proj1}
\eea
with
\begin{equation}
\mathcal{P} = \Gamma^{(2)}_N\Big|_{T=0} + R_N\,,
\label{proj2}
\end{equation}
and
\begin{equation}
\mathcal{F} = \Gamma^{(2)}_N - \Gamma^{(2)}_N\Big|_{T=0}\,.
\label{proj3}
\end{equation}
Since this expansion is essentially in powers of the fields/tensors and  our truncation involves up to hexic interaction terms, we are interested in the expansion \eqref{proj1} up to $n=3$. We evaluate the trace appearing in Eq.~\eqref{proj1}, which involves summing over all index variables and fields. The contraction of the Kronecker deltas in $\mathcal{P}^{-1}$ with open indices in $\mathcal{F}$ and subsequent trace over the appropriate power of $\mathcal{P}^{-1}\mathcal{F}$ gives rise to one of the invariants in Fig.~\ref{trunc} at the combinatorial level. However, an additional index dependence, arising from the regulator, also remains. Thence, the traces have the following structure,
\bea
&&{\text{Tr}}\mathcal{P}^{-1}(\partial_t R_N)\left(\mathcal{P}^{-1}\mathcal{F}\right)^n\nonumber\\
&=&\sum_I\sum_{a_i,b_j}\EuScript{F}(a_i,b_j)\times \mathrm{Invariant}^{(I)}\left(T^\#; a_k, b_l\right)\,,
\label{proj4}
\eea
where typically several invariants of Fig.~\ref{trunc}  involving $\#$ tensors are generated from a given order $n$.
The function $\EuScript{F}$  
depends on indices occurring in the contraction forming the invariant as well as on additional indices. As an example, consider a term in $\left(\mathcal{P}^{-1}\mathcal{F}\right)^2$ that contributes to $\beta_{{g}^{2,i}_{4,1}}$, shown in Fig.~\ref{fig:PFexpansion}.
The extra index-dependence of the invariant, i.e., the fact that crossed lines in Fig.~\ref{fig:PFexpansion} occur \textit{inside} the invariant, is a consequence of the $O(N')\otimes O(N')\otimes O(N')$ symmetry breaking introduced by the regulator. The flow generates terms that explicitly break that symmetry and feature the same combinatorial structures as the original invariants.

Accordingly, a full treatment requires a projection prescription that distinguishes symmetric and symmetry-breaking terms. We project out the flow of the symmetric terms by setting indices that occur in the invariant \textit{and} in the regulator to zero in the regulator. In other words, we first identify the invariant by the combinatorial structure. Second, we tag those indices that appear in the invariant and simultaneously in the regularized propagator. Those we set to zero in the regularized propagator. This differs from the projection prescription used in \cite{Eichhorn:2017xhy}, where the invariant was isolated and the sum over all indices in the regularized propagator was performed. 
 A concrete example that appears in the calculation is the following,
\begin{eqnarray}
&{}&{\rm Tr}(\mathcal{P}^{-1}\mathcal{F})^2|_{(g_{4,1}^0)^2}\nonumber\\
&\sim&\frac{1}{2}(g^{0}_{4,1})^2\sum_{\left\{a_i\right\},\left\{b_j\right\},c_1}\left(\frac{\partial_t Z_N + Z_N}{Z^3_N N }\right)(a_1+a_2+a_3)\nonumber\\
&{}&\times\frac{\theta\left(N-\sum_i a_i\right)}{1+\left(\frac{N}{a_1+b_2+b_3}-1\right)\theta\left(N-a_1-b_2-b_3\right)}\times\nonumber\\
&{}&\times T_{b_1 b_2 a_3}T_{b_1 a_2 b_3}T_{c_1 a_2 b_3}T_{c_1 b_2 a_3}\,.
\label{proj5}
\end{eqnarray}
In order to build the invariant, we need to perform the sum over $a_2,a_3,b_1,b_2,b_3,c_1$. The function $\EuScript{F}$ in this case is $\EuScript{F} = \EuScript{F}(a_1,a_2,a_3,b_2,b_3)$. Therefore, contracting the tensor indices and applying the projection rule, we obtain the following contribution,
\begin{eqnarray}
{\rm Tr}(\mathcal{P}^{-1}\mathcal{F})^2|_{(g_{4,1}^0)^2\, {\rm proj}}
&\sim&\frac{1}{2}(g^{0}_{4,1})^2\sum_{\left\{a_i\right\},\left\{b_j\right\},c_1}\EuScript{F}(a_1,0,0,0,0)\nonumber\\
&\times& T_{b_1 b_2 a_3}T_{b_1 a_2 b_3}T_{c_1 a_2 b_3}T_{c_1 b_2 a_3}\,.
\label{proj6}
\end{eqnarray}
One notices that the contributions removed by our projection rule are those in which the index present in $\EuScript{F}$ also appears as a contracted index in the tensors. These types of terms are called ``index-dependent" 
(in this classification, the regulator is an index-dependent term). Our projection rule allows for the computation of all spectral sums - or integrals in the large $N$ limit - as illustrated in Eq.~\eqref{proj6}. However, as soon as index-dependent interactions are taken into account in more sophisticated truncations, one needs to go beyond such a prescription.

\section{Search for universality classes}\label{sec:universalityclasses}
In this section we present the fixed-point results together with the associated critical exponents that we obtain using the FRG in a truncation up to sixth order in the tensors with a regulator of the form Eq.~\eqref{model6} and $r=1$. 
We first present the results for the corresponding beta functions, as these are quite illuminating in terms of the possible structure of fixed points. These are given for the dimensionless renormalized couplings, which are rescaled by an appropriate power of $N$ and $Z_N^{-n/2}$ for the $n$th power in tensors.

\subsection{Beta functions for $r=1$}\label{sec:betas}

\begin{itemize}
	\item Anomalous dimension: 
	The anomalous dimension is implicitly defined by
	\be
	\eta = \frac{1}{3}(4-\eta)(g^{2,1}_{4,1}+g^{2,2}_{4,1}+g^{2,3}_{4,1})+\frac{1}{10}(5-\eta)g^{2}_{4,2}\,,\label{eq:eta}
	\ee
	which can be solved leading to
	\be
	\eta = \frac{5\left( 8 \sum_{i=1}^3g_{4,1}^{2,i} + 3 g_{4,2}^2\right)}{30+10\sum_{j=1}^3 g_{4,1}^{2,j}+ 3 g_{4,2}^2 }.
	\ee
	\item Quartic couplings
	\begin{eqnarray}
	\beta_{{g}^{2,i}_{4,1}} &=&  (2+2\eta){g}^{2,i}_{4,1}+\frac{2}{5}(5-\eta)({g}^{2,i}_{4,1})^2+\frac{2}{3}(4-\eta)({g}^{0}_{4,1})^2\nonumber\\
	&-& \frac{1}{4}(4-\eta){g}^{3,i}_{6,1}-\frac{1}{40}(5-\eta){g}^{3,i}_{6,2}-\frac{1}{6}(4-\eta){g}^{2,i}_{6,1}\nonumber\\
	&-&\frac{1}{6}(3-\eta){g}^{0,p}_{6,1}-\frac{1}{2}(3-\eta){g}^{0,np}_{6,1}\,,
	\label{r1g2i41}
	\end{eqnarray}
	\begin{eqnarray}
	 {\beta}_{{g}^{2}_{4,2}}&=& (3+2\eta){g}^{2}_{4,2}+\frac{4}{5}(5-\eta){g}^{2}_{4,2}({g}^{2,1}_{4,1}+{g}^{2,2}_{4,1}+{g}^{2,3}_{4,1})\nonumber\\
	&+&\frac{4}{3}(4-\eta)({g}^{2,1}_{4,1}{g}^{2,2}_{4,1}+{g}^{2,1}_{4,1}{g}^{2,3}_{4,1}+{g}^{2,2}_{4,1}{g}^{2,3}_{4,1})\nonumber\\
	&+&\frac{2}{15}(6-\eta)({g}^{2}_{4,2})^{2}-\frac{1}{6}(3-\eta)({g}^{2,1}_{6,1}+{g}^{2,2}_{6,1}+{g}^{2,3}_{6,1})\nonumber\\
	&-& \frac{1}{6}(4-\eta)({g}^{3,1}_{6,2}+{g}^{3,2}_{6,2}+{g}^{3,3}_{6,2})-\frac{3}{40}(5-\eta){g}^{3}_{6,3}\,,\nonumber \\
	\label{r1g242}
	\end{eqnarray}
	\begin{eqnarray}
	{\beta}_{{g}^0_{4,1}}&=&\left(\frac{3}{2}+2\eta\right){g}^{0}_{4,1}-\frac{1}{12}(4-\eta)({g}^{1,1}_{6,1}+{g}^{1,2}_{6,1}+{g}^{1,3}_{6,1})\nonumber\\
	&-&\frac{1}{40}(5-\eta){g}^{1}_{6,2}\,.
	\label{r1g041}
	\end{eqnarray}
	One can infer several things about the fixed-point structure at the level of the quartic truncation from the structure of the beta function: Firstly, the symmetry under the interchange of colors in the cyclic melons implies that fixed points must either respect $g_{4,1}^{2,j\, \ast}= g_{4,1}^{2,i\, \ast}\, \forall i,j$, or come in triples that are mapped into each other under permutations of the $g_{4,2}^{2,i\, \ast}$. Secondly, one can consistently set $g_{4,1}^{0\, \ast}=0$, such that then $g_{4,1}^{2,2\, \ast}=0=g_{4,1}^{2,3\, \ast}$ which allows to set $g_{4,2}^{2, \ast}=0$. This provides a fixed point with only $g_{4,1}^{2,1\,\ast}\neq 0$ (and the corresponding two fixed points obtained by permuting the cyclic melonic couplings). This is a consequence of the enhanced $O(N)\otimes O(N^2)$ symmetry of the fixed point, which is reduced by the regulator to a $\mathbb{Z}_2$ symmetry mapping two of the indices into each other. The additional interactoin $g_{4,1}^0$ does not respect this symmetry.	\\
Further, one also infers the existence of a solution with $g_{4,1}^{2,1\,\ast}\neq 0$ and $g_{4,2}^{2\, \ast}\neq 0$. \\
	Note that one might at a first glance conclude that $g_{4,1}^{0\, \ast}=0$ at all fixed points. However, for $\eta = -3/4$, $\beta_{g_{4,1}^0}=0$ and then the nontrivial fixed-point value for $g_{4,1}^0$ is determined by demanding that $\beta_{g_{4,1}^{2,i}}=0$, see Sec.~\ref{qgcandidate}. 
	
	\item Sixth-order couplings
	\begin{eqnarray}
	{\beta}_{{g}^{3,i}_{6,1}}&=& (4+3\eta){g}^{3,i}_{6,1}+\frac{6}{5}(5-\eta){g}^{2,i}_{4,1}{g}^{3,i}_{6,1}+\frac{2}{3}(4-\eta){g}^{0}_{4,1}{g}^{1,i}_{6,1}\nonumber\\
	&-&\frac{16}{15}(6-\eta)({g}^{2,i}_{4,1})^3\,,
	\label{r1g3i61}
	\end{eqnarray}
	
	\begin{eqnarray}
	{\beta}_{{g}^{2,i}_{6,1}}&=& (4+3\eta){g}^{2,i}_{6,1}+\frac{1}{5}(5-\eta)\sum_{\substack{j,k=1 \\ j\neq i,k\, i\neq k}}^{3}{g}^{2,i}_{6,1}({g}^{2,j}_{4,1}+{g}^{2,k}_{4,1})\nonumber\\
	&+&\frac{2}{3}(4-\eta)\sum_{\substack{j,k=1 \\ j\neq i,k\, i\neq k}}^{3}{g}^{0}_{4,1}({g}^{1,j}_{6,1}+{g}^{1,k}_{6,1})\nonumber\\
	&+&(4-\eta)\sum_{\substack{j,k=1 \\ j\neq i,k\, i\neq k}}^{3} {g}^{0,np}_{6,1}({g}^{2,j}_{4,1}+{g}^{2,k}_{4,1})\nonumber\\
	&+&\frac{1}{3}(4-\eta)\sum_{\substack{j,k=1 \\ j\neq i,\, j\neq k\\i\neq k}}^{3} {g}^{0,p}_{6,1}({g}^{2,j}_{4,1}+{g}^{2,k}_{4,1})\nonumber\\
	&-& \frac{12}{5}(5-\eta)\sum_{\substack{j,k=1 \\ j\neq i,k\,i\neq k}}^{3}({g}^{0}_{4,1})^2({g}^{2,j}_{4,1}+{g}^{2,k}_{4,1})\,,
	\label{r1g2i61}
	\end{eqnarray}
	\begin{eqnarray}
	 {\beta}_{{g}^{1,i}_{6,1}}&=&\left(\frac{7}{2}+3\eta\right){g}^{1,i}_{6,1}+\frac{2}{5}(5-\eta){g}^{2,i}_{4,1}{g}^{1,i}_{6,1}+\frac{4}{3}{g}^{0}_{4,1}{g}^{0,p}_{6,1}(4-\eta)\,,\nonumber\\
	\label{r1g1i61}
	\end{eqnarray}
		\begin{eqnarray}
	{\beta}_{{g}^{1}_{6,2}}&=&\left(\frac{9}{2}+3\eta\right){g}^{1}_{6,2}+\frac{4}{5}(5-\eta){g}^{1}_{6,2}({g}^{2,1}_{4,1}+{g}^{2,2}_{4,1}+{g}^{2,3}_{4,1})\nonumber\\
	&+&\frac{2}{15}(6-\eta){g}^{2}_{4,2}{g}^{1}_{6,2}+\frac{2}{5}(5-\eta){g}^{2}_{4,2}({g}^{1,1}_{6,1}+{g}^{1,2}_{6,1}+{g}^{1,3}_{6,1})\nonumber\\
	&+&\frac{2}{3}(4-\eta){g}^{1,3}_{6,1}({g}^{2,1}_{4,1}+{g}^{2,2}_{4,1})+\frac{2}{3}(4-\eta){g}^{1,2}_{6,1}({g}^{2,1}_{4,1}+{g}^{2,3}_{4,1})\nonumber\\
	&+&\frac{2}{3}(4-\eta){g}^{1,1}_{6,1}({g}^{2,2}_{4,1}+{g}^{2,3}_{4,1})\,,
	\label{r1g162}
	\end{eqnarray}
	\begin{eqnarray}
	{\beta}_{{g}^{3}_{6,3}}&=&(6+3\eta){g}^{3}_{6,3}+\frac{6}{5}(5-\eta){g}^{3}_{6,3}({g}^{2,1}_{4,1}+{g}^{2,2}_{4,1}+{g}^{2,3}_{4,1})\nonumber\\
	&+& \frac{2}{3}(4-\eta){g}^{2}_{4,2}({g}^{2,1}_{6,1}+{g}^{2,2}_{6,1}+{g}^{2,3}_{6,1})+\frac{2}{5}(6-\eta){g}^{2}_{4,2}{g}^{3}_{6,3}\nonumber\\
	&+& \frac{4}{5}(5-\eta){g}^{2}_{4,2}({g}^{3,1}_{6,2}+{g}^{3,2}_{6,2}+{g}^{3,3}_{6,2})-\frac{8}{21}(7-\eta)({g}^{2}_{4,2})^3\nonumber\\
	&-&\frac{16}{5}(6-\eta)({g}^{2}_{4,2})^2 ({g}^{2,1}_{4,1}+{g}^{2,2}_{4,1}+{g}^{2,3}_{4,1})\nonumber\\
	&-&\frac{48}{5}(5-\eta){g}^{2}_{4,2}({g}^{2,1}_{4,1}{g}^{2,2}_{4,1}+{g}^{2,1}_{4,1}{g}^{2,3}_{4,1}+{g}^{2,2}_{4,1}{g}^{2,3}_{4,1})\,,
	\label{r1g363}
	\end{eqnarray}
	
	\begin{equation}
	{\beta}_{{g}^{0,np}_{6,1}} = (3+3\eta) {g}^{0,np}_{6,1}\,, 
	\label{r1g0np61}
	\end{equation}
	\begin{equation}
	{\beta}_{{g}^{0,p}_{6,1}} = (3+3\eta) {g}^{0,p}_{6,1}\,, 
	\label{r1g0p61}
	\end{equation}
	
	\begin{eqnarray}
	{\beta}_{{g}^{3,i}_{6,2}} &=& (5+3\eta){g}^{3,i}_{6,2}+\frac{6}{5}(5-\eta){g}^{2,i}_{4,1}{g}^{3,i}_{6,2}+\frac{6}{5}(5-\eta){g}^{2}_{4,2}{g}^{3,i}_{6,1}\nonumber
	\\&+&\frac{2}{15}(6-\eta){g}^{2}_{4,2}{g}^{3,i}_{6,2}+2(4-\eta){g}^{2}_{4,2}{g}^{0,np}_{6,1}+\frac{2}{3}(4-\eta)\times\nonumber\\
	&\times&{g}^{2}_{4,2}{g}^{0,p}_{6,1}+\frac{4}{3}(4-\eta){g}^{0}_{4,1}{g}^{1}_{6,2}-\frac{24}{5}(5-\eta){g}^{2}_{4,2}({g}^{0}_{4,1})^2\nonumber\\
	&-&\frac{16}{5}(6-\eta){g}^{2}_{4,2}({g}^{2,i}_{4,1})^2+\frac{2}{3}(4-\eta)\sum_{\substack{j=1 \\ j\neq i}}^{3}{g}^{2,j}_{4,1}{g}^{2,j}_{6,1}\nonumber\\
	&+& \sum_{\substack{j,k=1 \\ j\neq i,k,\,i\neq k}}^{3}\left(\frac{2}{3}(4-\eta){g}^{2,i}_{4,1}+\frac{(5-\eta)}{5}{g}^{2}_{4,2}\right)\left({g}^{2,j}_{6,1}+{g}^{2,k}_{6,1}\right)  \nonumber\\
	&+&\frac{1}{5}(5-\eta)\sum_{\substack{j,k=1 \\ j\neq i,k,\,i\neq k}}^{3}{g}^{3,i}_{6,2}({g}^{2,j}_{4,1}+{g}^{2,k}_{4,1})\nonumber\\
	&+&(4-\eta)\sum_{\substack{j,k=1 \\ j\neq i,k,\,i\neq k}}^{3}{g}^{3,i}_{6,1}({g}^{2,j}_{4,1}+{g}^{2,k}_{4,1})\nonumber\\
	&-&\frac{12}{5}(5-\eta)\sum_{\substack{j,k=1 \\ j\neq i,k,\, i\neq k}}^{3}({g}^{2,i}_{4,1})^2 ({g}^{2,j}_{4,1}+{g}^{2,3}_{4,1})\,.
	\label{r1g3i62}
	\end{eqnarray}
\end{itemize}

\subsection{Dynamical decoupling of multitrace interactions}

Imposing $O(N^\prime)\otimes O(N^\prime) \otimes O(N^\prime)$ invariance on the interaction terms does not forbid the construction of ``multitrace" invariants, such as, e.g., $T^2 T^2=T_{abc}T_{abc}T_{def}T_{def}$.  Diagrammatically, those interactions are represented by disconnected graphs. However one should keep in mind that they represent vertices and not Feynman diagrams, see Fig.~\ref{trunc}. Accordingly, they are somewhat analogous to non-local interactions in QFTs on a background. In the present background-independent setting such ``non-localities" are unavoidable.
In particular, the  ``single-trace" invariants are already combinatorically non-local, i.e, the theory space is not spanned by combinatorically local interactions. Consequently, working with the one-particle irreducible generating functional does not exclude ``multitrace" invariants. In particular, if one chooses a truncation for the effective average action $\Gamma_N$ which does not involve ``multitrace" (or ``disconnected") interactions, they are typically generated along the RG flow as a direct consequence of the combinatorial non-trivial nature of the vertices of tensor models.
This is in contrast to standard quantum field theories based on local vertices. There, although non-localities are present in the effective action, they arise by the resummation of local terms and truncations containing local vertices will not generate non-local ones. 

We stress that multitrace interactions contribute to the anomalous dimension and beta functions of single-trace couplings and could even generate new relevant directions, see \cite{Eichhorn:2017xhy} for the rank-3 complex tensor model case and \cite{BenGeloun:2018ekd} for group field theories. Accordingly, they should not simply be neglected in truncations. Further, there is no symmetry that allows the multitrace interactions to be consistently set to zero, to the best of our knowledge. In fact, the opposite appears to be true, as multitrace interactions of the form $(T_{abc}T_{abc})^n$ actually feature an \emph{enhanced} symmetry $O(N'^3)$. Therefore, it is even consistent to consider a theory space where \emph{only} these interactions exist, but any theory space that features single-trace interactions with $O(N')\otimes O(N')\otimes O(N')$ symmetry necessarily contains the $O(N'^3)$ symmetric theory space as a subspace.
As these interactions are generated, there is not only an upper but also a lower bound on their canonical scaling dimension. These two agree (if the quartic cylcic melonic interactions in turn satisfy their upper bound). Accordingly, it is not possible to decouple the multitrace interactions by simply choosing a canonical dimension below their upper bound.

From the geometric point of view, multitrace interactions  lead to surfaces with disconnected boundaries and might naively be associated with ``baby universes". Hence, one might tentatively conjecture that a suitable continuum limit requires a dynamical decoupling of such interactions. Dynamical decoupling refers to a fixed-point structure at which these interactions vanish. The beta functions of multitrace couplings contain terms which are finite even if all multitrace couplings are set to zero. Accordingly, we search for a fixed-point structure where these additional contributions vanish at the fixed point.
 In \cite{Eichhorn:2017xhy}, it was discussed how to obtain such a decoupling by \textit{not} imposing an additional symmetry between the three distinct colors. The symmetry would lead to $g^{2,1}_{4,1}=g^{2,2}_{4,1}=g^{2,3}_{4,1}$ etc.. In order to show this in a concrete example, let us consider the beta functions for the rank-3 real tensor model in a simple truncation including interactions up to quartic interaction terms, i.e., the effective action is given by Eqs.~\eqref{model3} and \eqref{model4}. The beta functions are shown in Eq.~\eqref{eq:eta}, Eq.~\eqref{r1g2i41}, Eq.~\eqref{r1g242} and Eq.~\eqref{r1g041}.
By assumption, we are looking for fixed points such that $g^{2\ast}_{4,2} = 0$. From Eq.~\eqref{r1g242}, one obtains 
\begin{equation}
(4-\eta)(g^{2,1}_{4,1}g^{2,2}_{4,1}+g^{2,1}_{4,1}g^{2,3}_{4,1}+g^{2,2}_{4,1}g^{2,3}_{4,1}) = 0\,.
\label{decoupl4}
\end{equation}
Eq.~\eqref{decoupl4} is satisfied either by setting $\eta = 4$ which violates regulator bounds (see \cite{Meibohm:2015twa} and Sec.~\ref{sec:FPcriteria} for details) or 
\begin{equation}
g^{2,1}_{4,1}g^{2,2}_{4,1}+g^{2,1}_{4,1}g^{2,3}_{4,1}+g^{2,2}_{4,1}g^{2,3}_{4,1} = 0\,.
\label{decoupl5}
\end{equation}
Accordingly, to allow for $g_{4,2\, \ast}^2=0$ at a non-trivial fixed point, only one of the cyclic melonic couplings may be finite. Inspecting  Eq.~\eqref{r1g041}, we observe that $g_{4,1\, \ast}^0=0$ is a fixed point. At this fixed point, all contributions in $\beta_{g_{4,1}^{2,i}}$ are proportional to $g_{4,1}^{2,i}$. Thence, we can choose, e.g., $g_{4,1\, \ast}^{2,1}=0=g_{4,1\,\ast}^{2,2}$ and obtain a fixed point at which $g_{4,1\, \ast}^{2,3}=\frac{5+5\eta}{5-\eta}$. The multitrace interaction dynamically decouples at this fixed point.

The same feature was observed in the complex rank-3 tensor model, see \cite{Eichhorn:2017xhy}. One realizes that if cyclic melonic interactions were not introduced with distinct couplings, it would be impossible to find a fixed-point with decoupled multitrace interactions. This agrees with the results reported in \cite{BenGeloun:2018ekd}. In summary, a continuum limit with ``single-trace" interactions necessarily requires anisotropic couplings for the melonic interactions.

\subsection{Approximations for the beta functions and fixed-point searches}\label{sec:FPcriteria}
At any order in the truncation, zeros of the beta functions exist which are truncation artefacts. With the 21 couplings in our truncation, the total number of zeros is quite large.
To distinguish these from fixed points which could be physical, we impose a set of conditions.
These include 
\begin{itemize}
\item[a)] continued existence under extensions of the truncation starting with the quartic truncation,
\item[b)] respecting the regulator bound $\eta< r$,
\item[c)] critical exponents that are compatible with our truncation scheme according to canonical dimension.
\end{itemize}

For the first condition, we  explore three different approximations that differ in their treatment of the anomalous dimension. Firstly, setting $\eta=0$ defines the simplest scheme, analogous to the local potential approximation in local quantum field theory. Secondly, setting $\eta=0$ everywhere where it arises from the scale-derivative of the regulator defines the perturbative approximation. Its name derives from the fact that the universal one-loop beta functions for dimensionless couplings in local quantum field theories are recovered from the FRG under this approximation. Note that only the ``canonical" $\eta$-terms, arising from the definition of the renormalized couplings, are thereby taken into account. Finally, taking the full, non-polynomial expression for $\eta$ into account is an approximation that might be considered as resumming higher orders in the couplings.\\
Traceability under extensions of the truncation consists in the following requirement: Going from $\eta=0$ in quartic order to the full approximation at sixth order, at each of the 5 steps of the extension of the truncation, the variation in critical exponents should not be larger than $\mathcal{O}(1)$. Therefore, the number of relevant/irrelevant directions should only change between truncations if $|\theta_i|<1$. Fixed-point values and critical exponents should also vary less and less, the further the truncation is extended. Zeros of the beta functions which only occur beyond a certain order in the truncation are not considered as candidates for actual fixed points.

The consistency of the regulator imposes the bound $\eta<r$, derived as follows \cite{Meibohm:2015twa}.
We choose a regulator that is proportional to the wave-function renormalization $Z_N$. In turn, its scale-derivative is related to the anomalous dimension	
  $\eta=-\partial_t Z_N/Z_N$. This definition can be rewritten as follows
		\begin{equation}
		Z_N\sim N^{-\eta}.
		\label{ZScaling}
		\end{equation}
In the limit where $N\rightarrow N'\rightarrow\infty$, the effective action $\Gamma_N$ should converge to the classical action $S$. This requires that the regulator diverges in this limit,
		\begin{equation}
		\lim\limits_{N\rightarrow N'\rightarrow \infty}R_N\rightarrow \infty.
		\label{RegulatorConstraint}
		\end{equation}
	Using our choice of  shape function for general $r$, we have the following scaling-behaviour of the regulator as $N\rightarrow\infty$
		\begin{equation}
		\lim\limits_{N\rightarrow N'\rightarrow \infty}R_N\sim Z_N\,N^r.
		\label{eq:regScaling}
		\end{equation}
Inserting the $N$-dependence of the wave-function renormalization from Eq.~\eqref{ZScaling}, we obtain
		\begin{equation}
		R_N\sim N^{r-\eta}.
		\label{eq:regScaling2}
		\end{equation}
	Thus,
		 the anomalous dimension must satisfy the following inequality 
		\begin{equation}
		\eta <  r,
		\end{equation}
		in order to satisfy the constraint given by Eq.~\eqref{RegulatorConstraint}.
Zeros of the beta function which violate this condition are not consistent fixed-point candidates with our choice of regulator. This does not imply that fixed points cannot feature larger anomalous dimensions. However, in that case one should choose a regulator which is independent of the wave-function renormalization to re-investigate such zeros.

Finally, we implement the third condition by demanding that the maximum shift $\theta_1-d_{\bar{g}}$ between the largest critical exponent and the largest canonical dimension  (which is $-3/2$ in the case of $r=1$) should not be larger than 5. Even under the assumption that all scaling dimensions at the fixed point are shifted by such a large amount, no couplings beyond our truncation can become relevant. Accordingly, fixed-point candidates with critical exponents slightly larger than 3 definitely require further investigation. Zeros of the beta functions with critical exponents larger than $\theta=3.5$ are not included in our list of fixed-point candidates, so we require
\be
\theta \leq 3.5.\label{eq:bound}
\ee
Note that this does not preclude the existence of fixed points with larger critical exponents. It only implies that to reliable establish their existence, larger truncations have to be explored and/or a different ordering principle than the near-canonical guiding has to be demanded.
 One can also ``symmetrize" the bound (\ref{eq:bound}) in the sense of demanding that the deviation of the smallest critical exponent from the smallest canonical dimension should be bounded.

At sixth order in the truncation, the number of couplings is 21, making the fixed-point search slightly more challenging than in the quartic truncation, where even the perturbative truncation can be solved quickly and easily, using, e.g., Mathematica's NSolve. At higher order in the truncation, it is therefore useful to impose symmetry-constraints to reduce the number of beta functions. For instance, when searching for iscolored fixed points, i.e., those which are symmetric under permutations of the colors, a reduced set of only 11 beta functions needs to be solved, with all other fixed-point values determined by symmetry.\\

 The fixed-point candidates we discover are the following: 
 \begin{itemize}
 \item Fixed points featuring dimensional reduction
\begin{itemize}
	\item a cyclic-melonic single-trace fixed point: only, e.g., $g^{2,1}_{4,1}$ and $g^{3,1}_{6,1}$ are non-zero (plus their color permutation),
	\item a multitrace-bubble fixed point: only the disconnected bubbles associated to maximum number of disconnected components $n/2$ at each order $n$ in the number of tensors, i.e., $g^{2}_{4,2}$ and $g^{3}_{6,3}$ do not vanish,
	\item a cyclic-melonic multitrace fixed point: only, the cylic melonic interactions with arbitrarily many traces e.g., $g^{2,1}_{4,1}$, $g^{2}_{4,2}$, $g^{3,1}_{6,1}$, $g^{3,1}_{6,2}$ and $g^{3}_{6,3}$ are non-trivial (and the color permutations),
	\end{itemize}
\item Fixed-point candidates for three-dimensional quantum gravity	
	\begin{itemize}
	\item an isocolored fixed point with tetrahedral interaction: only $g^{0}_{4,1}$, $g^{2,i}_{4,1}$, $g^{2}_{4,2}$, $g^{2,i}_{6,1}$, $g^{3,i}_{6,1}$, $g^{1}_{6,2}$, $g^{3,i}_{6,2}$ and $g^{3}_{6,3}$, $\forall i$ are non-zero,
	\item an isocoloured fixed point without geometric interaction: only $g^{2,i}_{4,1}$, $g^{2}_{4,2}$, $g^{3,i}_{6,1}$, $g^{3,i}_{6,2}$ and $g^{3}_{6,3}$ are non-vanishing. This fixed point is shifted into the complex plane and accordingly only discussed in the appendix.
\end{itemize}
\end{itemize}

\subsection{Indications of stability of the potential}\label{sec:potstab}

As an additional requirement on viable fixed points, one might impose that there are no unstable directions of the action in configuration space. 
A comprehensive study in configuration space and beyond polynomial truncations is beyond the scope of this work. Instead we limit ourselves to selecting a simple field configuration and investigating the form of the action for this choice of configuration within our truncation.
 Accordingly we consider the following configuration
	\begin{equation}
	T_{a_1 a_2 a_3}=t\,\delta_{a_1 a_2},
	\end{equation}
	 $\forall \, a_3 $ and  $t \in \mathbb{R}$. For this configuration, the quartic multitrace interaction evaluates to
	\begin{equation}
	T_{a_1 a_2 a_3}T_{a_1 a_2 a_3}T_{b_1 b_2 b_3}T_{b_1 b_2 b_3}= t^4\, N^4.
	\end{equation}
	Furthermore, for the cyclic melons we obtain
	\bea
	T_{a_1 a_2 a_3}T_{a1 a_2 b_3}T_{b_1b_2b_3}T_{b_1b_2a_3}&=& t^4\, N^4,\label{eq:tg4121}\\
	T_{a_1 a_2a_3}T_{a_1b_2a_3}T_{b_1b_2b_3}T_{b_1a_2b_3}&=&t^4\, N^3,\label{eq:tg4122}\\
	T_{a_1a_2a_3}T_{a_1 b_2b_3}T_{b_1b_2b_3}T_{b_1a_2a_3}&=&t^4\, N^3.\label{eq:tg4123}
	\eea
The difference in scaling between Eq.~\eqref{eq:tg4121} and Eqs.~\eqref{eq:tg4122}, \eqref{eq:tg4123} is due to the choice $T_{a_1a_2a_3}\sim \delta_{a_1 a_2}$.
	The quartic effective potential then takes the following form
	\begin{eqnarray}
	V(T)&=&
	\bar{g}^{2,1}_{4,1}T_{a_1 a_2 a_3}T_{b_1 a_2 a_3}T_{a_1 b_2 b_3}T_{b_1 b_2 b_3}\nonumber\\&+&\bar{g}^{2,1}_{4,1}T_{a_1 a_2 a_3}T_{a_1 b_2 a_3}T_{b_1 a_2 b_3}T_{b_1 b_2 b_3}\nonumber\\
	&+&\bar{g}^{2,1}_{4,1}T_{a_1 a_2 a_3}T_{a_1 a_2 b_3}T_{b_1 b_2 a_3}T_{b_1 b_2 b_3}\nonumber\\
	&+&\bar{g}^0_{4,1}T_{a_1 a_2 a_3}T_{a_1 b_2 b_3}T_{b_1 a_2 b_3}T_{b_1 b_2 a_3}\nonumber\\
	&+&\bar{g}^2_{4,2}T_{a_1 a_2 a_3}T_{a_1 a_2 a_3}T_{b_1 b_2 b_3}T_{b_1 b_2 b_3}\nonumber\\[2mm]
	&=& \left(g^{2,1}_{4,1}+g^{2,2}_{4,1}\right)N^{-2} N^3\,t^4+g^{2,3}_{4,1} N^{-2}N^4\,t^4\nonumber\\
	&+& g^0_{4,1}N^{-3/2} N^3\, t^4+g^2_{4,2} N^{-3}N^4\,t^4.
	\end{eqnarray} 
	At large $N$, the interaction parameterized by $g_{4,1}^{2,3}$ dominates. Within the quartic truncation one would accordingly demand $g_{4,1\ast}^{2,3}>0$ for the action in this truncation to be bounded from below at the fixed point.  Given that there are nonzero terms beyond the truncation, this conclusion is premature.
	If we also include sixth order interactions, we find that in the large-$N$ limit the following terms dominate in the potential:
	\begin{equation}
	V(t)=g^{2,3}_{4,1}N^2\,t^4+g^{3,3}_{6,1}N^2\,t^6.
	\end{equation}
	The stability criterion is accordingly pushed from the quartic to the hexic coupling. Under further extensions of the truncation, this persists. Yet, one should keep in mind that instead of performing a polynomial expansion one could also evaluate the fixed-point equation for the full potential (e.g., in the cyclic melonic single-trace interactions). The polynomial expansion is expected to approximate the full solution within a finite radius of convergence. There are then several possibilities, including: a) The potential is bounded from below with the global minimum at $T=0$. In this case one would expect all Taylor coefficients of the expansion to be positive. b) The potential is unbounded from below with a maximum at $T=0$. In this case all Taylor coefficients are expected to be negative. c) The potential is bounded from below with a minimum away from $T=0$. This corresponds to a case of spontaneous symmetry breaking, e.g., of the $\mathbb{Z}_2$ reflection symmetry under which $T \rightarrow -T$. In this case the low-order Taylor coefficients in an expansion around $T=0$, as in our truncation, are expected to be negative, with the higher-order ones becoming positive. Alternatively, the Taylor coefficients can also alternate between positive and negative, such that the potential alternates between a seemingly unstable and stable form at large values of the field, outside the radius of convergence.\\
Accordingly, while the analysis of the potential in a polynomial truncation cannot provide full information on the stability of the potential, it can give hints. In this spirit, we will come back to the three cases discussed above.

\section{Universality classes with dimensional reduction}\label{uniclassdimred}
\subsection{Cyclic-Melonic Single-trace Fixed Point}
At the cyclic melonic single-trace fixed point (CMSTFP) only one of the couplings of the cyclic 
melonic single trace interactions
takes a non-trivial value, while all other 
coupling
vanish.  
This implies an enhanced symmetry, where one passes from an  $O(N')\otimes O(N')\otimes O(N')$-symmetric model to a $O(N')\otimes O(N'^2)$-symmetric one. This enhanced symmetry is not obeyed by the melonic interactions which are not cyclic, which explains their decoupling at this fixed point\footnote{We thank D.~Benedetti for pointing this out to us.}. The set of critical exponents reflects this symmetry as there is a pair of degenerate critical exponents already in the quartic truncation, associated to the approach to the fixed point along two eigendirections that can be mapped into each other by interchanging two colors.  Indeed, it can be easily checked that at this fixed point at quartic order the stability matrix is diagonal in  $g^{2,2}_{4,1}$ and $g^{2,3}_{4,1}$ if $g^{2,1\ast}_{4,1}$ does not vanish. Specifically,  the entries of the two rows $-\frac{\partial \beta_{{g}^{2,2}_{4,1}}}{\partial g_i}$ and $-\frac{\partial \beta_{{g}^{2,3}_{4,1}}}{\partial g_i}$, where $g_i$ is the vector of all couplings, only contain non-zero components for $g=g^{2,2}_{4,1}$ and $g=g^{2,3}_{4,1}$, respectively. Furthermore, it can be shown that at the cyclic melonic fixed point $-\frac{\partial \beta_{{g}^{2,2}_{4,1}}}{\partial {{g}^{2,2}_{4,1}}}=-\frac{\partial \beta_{{g}^{2,3}_{4,1}}}{\partial {{g}^{2,3}_{4,1}}}$. This explains the existence of a pair of degenerate critical exponents. The same reasoning explains the degeneracies at hexic order.

The same fixed-point structure was already observed in the complex tensor model, \cite{Eichhorn:2017xhy}. As the difference between the complex and the real model is the existence of the interactions $g_{4,1}^0$ and $g_{6,1}^{0,p}$  which can consistently be set to zero if only the cyclic melonic couplings with one preferred color are nonvanishing, the existence of this fixed-point structure in both models is expected. Comparing (the non-universal) fixed-point values to those in \cite{Eichhorn:2017xhy} we observe a substantial difference. On the other hand, the universal critical exponents take very similar values for the relevant direction ($\theta=2.21$ for the complex and $\theta=2.19$ for the real case, see  Tables~\ref{cycmelsttable1} and \ref{cycmelsttable2}),  albeit different projection rules for the FRG equation were employed. Further, the least negative irrelevant critical exponents,  see Table~\ref{cycmelsttable2}, are also very similar, and allow us to identify $\theta_2=-0.03$ as a new direction in the real model, associated to the existence of the geometric interaction $g_{4,1}^0$. We stress that while the value of $\theta_2$ is negative, the systematic error of our truncation is expected to be significantly larger than the deviation of $\theta_2$ from zero. Accordingly, we cannot exclude that the new interaction $g_{4,1}^0$ might in fact be associated to a relevant direction. Extended truncations are required to explore this question. 

According to the discussion in Sec.~\ref{sec:potstab}, the negative sign of both nonzero interactions could indicate an unstable potential. This is a first property this fixed point has in common with the double-scaling limit in matrix models  \cite{Douglas:1989ve,Brezin:1990rb,Gross:1989vs,Gross:1989aw}.

The symmetry-structure of the fixed point has the following implication: Assuming for instance that $1$ is the preferred colour, i.e., only $\dg2141=\dg3161\neq 0$, then said colour $1$ as well as the pair ${23}$ are invariant under the $O(N^2)$ group. 
In fact, the fixed-point structure is such that the index
pair ${23}$ can be the understood as a single 
index. In other words, the combinatorics of the graphs generated by this fixed-point action agree with those of a matrix model. 
One might view this as a phenomenon of dimensional reduction: Starting with three-dimensional building blocks, a universal continuum limit could exist where the emergent physics is that of two-dimensional gravity.
Yet, the leading critical exponent differs from that of matrix models for two-dimensional quantum gravity (which is $\theta_m=0.8$), at least in our truncation for $r=1$ (see Eq.~\eqref{model6}). This is a consequence of the different canonical scaling dimensions of $g_{4,1}^{2,1}$ in comparison to the quartic interaction in a matrix model, which has canonical dimension -1: In the simplest approximation of the beta functions of the quartic interactions,
\be
\beta_{g_4} = -d_{\bar{g}_4}g_4+ \# g_4^2,
\ee
leads to a fixed point at $g_{4\, \ast} = \frac{d_{\bar{g}_4}}{\#}$ with critical exponent $\theta = -d_{\bar{g}_4}$. For the 
matrix model, this leads to the estimate $\theta=1$ (see, e.g., \cite{Eichhorn:2013isa}).
Inspecting Eq.~\eqref{eq:lowerbounds} for the upper bounds on scaling dimensions in the tensor model, we find that choosing $r=1/2$ (which is a viable choice satisfying all regulator requirements), results in $\theta=1$ also for the tensor model in this approximation. Further, this is the choice that yields $[g_{4,1}^{2,1}]=-1$, as it should be for the canonical dimension of the quartic coupling in the matrix model.\\
Accordingly, we conjecture that the universality class of two-dimensional quantum gravity can be recovered from tensor models at the cyclic melonic fixed point for the choice $r=1/2$. This observation is strengthened by more sophisticated approximations of the beta functions: for the quartic truncation with the full effect of the anomalous dimension, a fixed point with one relevant direction $\theta =1.13$ is obtained. In the hexic truncation, the same structure remains and the relevant critical exponent is $\theta =1.09$. These values are quantitatively similar to the results obtained with the FRG for the double-scaling limit in matrix models for 2d quantum gravity, see \cite{Eichhorn:2013isa}. That tensor models feature a universal continuum limit in agreement with two-dimensional quantum gravity is in fact very encouraging: It implies that our implementation of universality in the sense of an RG fixed point in the ``scale" $N$ is indeed viable.

\begin{widetext}
	
	\begin{table}[H]
		\begin{tabular}{ |c||c|c|c|c|c|c| } 
			\hline
			scheme & ${g_{4,1}^{0}}^*$ & ${g_{4,1}^{2,1}}^*$ & ${g_{4,1}^{2,2}}^*$ & ${g_{4,1}^{2,3}}^*$  & ${g_{4,2}^{2}}^*$ &\\ \hline 
			full & 0 & -0.38 & 0 & 0 & 0&  \\ 
			pert & 0 & -0.43 & 0 &0&0&\\
			$\eta=0$ & 0 & -1 & 0 & 0 & 0& \\ 
			\hline
			\hline
			scheme & $\theta_1$ & $\theta_2$ & $\theta_3$ & $\theta_4$ & $\theta_5$ & $\eta$\\ \hline 
			 full & 2.27 &   
			-0.16 & -0.34& -0.84 &  -0.84 & -0.58  \\ 
			 pert & 2 & -0.14 & -0.36 &  -0.86 & -0.86 & -0.57 \\
			$\eta=0$ &  
			2 & 1 & -1.5& -2 & -2 & -\\ 
			\hline
		\end{tabular}\\
		\caption{
			The  cyclic-melonic single-trace fixed point features one relevant direction in the quartic truncation.
		}
	\label{cycmelsttable1}
	\end{table}

	\begin{table}[H]
		\begin{tabular}{ |c||c|c|c|c|c|c|c|c|c|c|c|c|c|c|c|c| } 
			\hline
			scheme & ${g_{4,1}^{0}}^*$ & ${g_{4,1}^{2,1}}^*$ & ${g_{4,1}^{2,2}}^*$ & ${g_{4,1}^{2,3}}^*$& ${g_{4,2}^{2}}^*$& ${g^{0,np}_{6,1}}^*$ & ${g^{0,p}_{6,1}}^*$ & ${g^{1,i}_{6,1}}^*$ & ${g^{2,i}_{6,1}}^*$  & ${g^{3,1}_{6,1}}^*$ & ${g^{3,2}_{6,1}}^*$ & ${g^{3,3}_{6,1}}^*$& ${g^{1}_{6,2}}^*$ & ${g^{3,i}_{6,2}}^*$  & ${g^{3}_{6,3}}^*$   \\ \hline 
			full & 0 & -0.28 & 0 & 0 & 0 & 0 & 0 & 0 & 0  & -0.15 & 0 & 0 & 0  & 0 & 0   \\ 
			pert&0&-0.30&0 & 0 & 0& 0& 0 & 0 & 0 &-0.18&  0 & 0 & 0 & 0 & 0    \\
			$\eta=0$  & 0 & -0.46  & 0 & 0 & 0& 0 & 0 & 0 & 0& -0.50 & 0  & 0 & 0 & 0  & 0  \\
			\hline
			\hline
			scheme & $\theta_1$ & $\theta_2$ & $\theta_3$ & $\theta_{4,5}$  & $\theta_6$ & $\theta_{7,8}$ & $\theta_9$ & $\theta_{10}$ & $\theta_{11,12}$  & $\theta_{13,14}$  & $\theta_{15,16,17}$ & $\theta_{18}$ & $\theta_{19}$ & $\theta_{20,21}$  & $\eta$\\ \hline 
			full & 2.19 & -0.03 &  -0.69 &  -1.19  & -1.68 & -1.78 & -2.08 & -2.15 & -2.18 & -2.28 & -2.78  & -2.94 & -2.98 & -3.18 &-0.41 \\  
			pert &2& -0.03&-0.70 & -1.20& -1.69 &-1.79&-2.09&-2.17&-2.19&-2.29&-2.79&-2.94&-2.98&-3.19&-0.40\\
			 $\eta=0$  & 2 & 0.53 & -1.5 & -2 & -2.58   & -3& -2.66 & -3.25  & -3.08 & -3.5 & -4& -3.41& -3.94 & -4.08 &-\\ 
			\hline
		\end{tabular}\\
		\caption{The cyclic-melonic single-trace fixed point only features cyclic melonic couplings and one associated relevant direction in the truncation to sixth order. Given the systematic error of the truncation and the small value $\theta_2=-0.03$, a second positive critical exponent might exist at higher orders in the truncation.}
	\label{cycmelsttable2}
	\end{table}
\end{widetext}

\subsection{Multitrace-bubble Fixed Point}
At the multitrace bubble fixed point merely the melonic multitrace sector with $n/2$ ``traces" at order $n$ in the tensors is interacting (i.e., only couplings of the type $g^n_{2n,n}$ are non-vanishing). These interactions are the ``bubbles".
This fixed point has also been found in the complex model 
\cite{Eichhorn:2017xhy} with 
one relevant direction. 
The critical exponents associated to this fixed point agree with the ones found in the real model at hand. Due to an enlarged theory space, however, a second relevant direction $\theta=0.18$ is picked up in the quartic truncation, see Table~\ref{multitracebubbletable1}. We caution that the systematic error of a quartic truncation is expected to be significantly larger than 0.18 in the critical exponents. Therefore, this truncation is insufficient to determine the number of relevant directions. In fact, the truncation to order $T^6$ yields a negative critical exponent, see Table~\ref{multitracebubbletable2}. It is straightforward to extend to a $T^8$ truncation for the multitrace-bubble fixed point if one takes into account that only bubbly diagrams take non-trivial fixed point values. Therefore, in order to determine all fixed point values in a full $T^8$ truncation, one only needs to consider the contributions of $g^4_{8,4}$ to $\beta_{g^3_{6,3}}$ as well as the contributions to $\beta_{g^4_{8,4}}$ coming from $g^2_{4,2}$ and $g^3_{6,3}$.  The running of $g^3_{6,3}$ then has the following additional term
\begin{equation}
\beta_{g^3_{6,3}}\Big|_{g_{8,4}^4}=-\frac{(5-\eta)}{5} g^4_{8,4},
\end{equation}
while the beta function of $g^4_{8,4}$ at the multitrace melonic fixed point is given by
\begin{eqnarray}
	{\beta}_{{g}^{4}_{8,4}}&=& (9+4\eta){g}^{4}_{8,4}\,+\frac{8}{7}(8-\eta)(g^2_{4,2})^4\nonumber\\
	&&- \frac{12}{7}(7-\eta)g^3_{6,3}(g^2_{4,2})^2\nonumber\\
	&&+\frac{3}{10}(6-\eta)(g^3_{6,3})^2.
\end{eqnarray}
Note that the above terms included in the running of $g^4_{8,4}$ are the only ones that will contribute to the stability matrix $\mathcal{M}_{ij}=\partial \beta_i/\partial {g_j}$. Taking into account ${\beta}_{{g}^{4}_{8,4}}$, we find that the critical exponents remain stable under this extensions of truncation, still featuring one relevant directions, as in the hexic truncation. Indeed, for $\eta=0$, $\theta_1=3$, while $\theta_2=-1.5$. When increasing the truncation by a perturbative approximation, we find that $\theta_1=3$, while $\theta_2=-0.65$. Finally, in the full non-perturbative approximation $\theta_1=3.24$ while $\theta=-0.64$. The anomalous dimension is given by $\eta_{\text{full}}=-0.43$.

The fixed point features an enhanced $O(N^3)$ symmetry. Accordingly the three indices can be summarized to one ``superindex", meaning that it reduces to a vector model. This is an even more significant dimensional reduction that for the CMSTFP.
\\

	\begin{table}[H]
		\begin{tabular}{ |c||c|c|c|c|c|c| } 
			\hline
			scheme & ${g_{4,1}^{0}}^*$ & ${g_{4,1}^{2,1}}^*$ & ${g_{4,1}^{2,2}}^*$ & ${g_{4,1}^{2,3}}^*$  & ${g_{4,2}^{2}}^*$&\\ \hline 
			full & 0 & 0 & 0 & 0 & -1.44 &  \\ 
			pert & 0 &0 &0&0 & -1.67 &\\
			$\eta=0$  & 0 & 0 & 0 & 0 & -3.75& \\ 
			\hline
			\hline
			scheme & $\theta_1$ & $\theta_2$ & $\theta_3$ & $\theta_4$ & $\theta_5$ & $\eta$\\ \hline 
			full & 3.47 & 0.18 &  -0.32 &  -0.32 & -0.32 & -0.84\\ 
			 pert & 3 & 0.17 &   -0.33 &   -0.33 &   -0.33 & -0.83 \\
			($\eta=0$) & 3 & -1.5 & -2 & -2 & -2 &- \\ 
			\hline
		\end{tabular}\\
		\caption{
			The multitrace-bubble fixed point features two relevant directions in the truncation to quartic order, with one of the positive critical exponents rather close to zero, calling for an extension of the truncation to confirm its relevance.
		}
	\label{multitracebubbletable1}
	\end{table}
\begin{widetext}	
	
	\begin{table}[H]
		\begin{tabular}{ |c||c|c|c|c|c|c|c|c|c|c|c|c|} 
			\hline
			sch. & ${g_{4,1}^{0}}^*$ & ${g_{4,1}^{2,i}}^*$ & ${g_{4,2}^{2}}^*$& ${g^{0,np}_{6,1}}^*$ & ${g^{0,p}_{6,1}}^*$ & ${g^{1,i}_{6,1}}^*$  & ${g^{2,i}_{6,1}}^*$   & ${g^{3,i}_{6,1}}^*$  & ${g^{3,i}_{6,2}}^*$ & ${g^{1}_{6,2}}^*$  & ${g^{3}_{6,3}}^*$   \\ \hline 
			full        & 0 & 0 & -1.05 & 0  & 0 & 0 & 0 & 0& 0    & 0 & -2.27   \\ 
						pert & 0 & 0& -1.16 & 0& 0 & 0 & 0 & 0  & 0& 0  & -2.82   \\
			$\eta=0$   & 0  & 0 & -1.73 & 0& 0 & 0  & 0   & 0 & 0& 0  & -7.45  \\ 
			\hline
		\end{tabular}\\
		\begin{tabular}{ |c||c|c|c|c|c|c|c|c|c|c|c| }
			\hline
			sch. & $\theta_1$ & $\theta_2$ & $\theta_{3,4,5}$ &  $\theta_{6,7}$  & $\theta_{8,9,10}$  & $\theta_{11}$ & $\theta_{12,13,14,15,16,17}$ & $\theta_{18,19,20}$  & $\theta_{21}$& $\eta$\\ \hline 
			full & 3.32 & -0.33 & -0.83 &  -1.24  & -1.74 & -1.82 & -2.24 & -2.32  & -3.28&-0.59 \\ 
			pert& 3 & -0.34 & -0.84  & -1.26 & -1.76  & -1.83 & -2.26 & -2.33& -3.30 & -0.58\\
			$\eta=0$  & 3 & -1.5 & -2 & -3  & -3.5  & -3.12  & -4 & -3.62  & -5.08&-\\ 
			\hline
		\end{tabular}\\
		\caption{
			Results for the  multitrace-bubble fixed point at order $T^6$ in the truncation.  The leading critical exponent is close to our bound $\theta<3.5$, yet the explicit extension to $T^8$ shows the fixed point persists.}
	\label{multitracebubbletable2}
	\end{table}
\end{widetext}

\subsection{Cyclic-Melonic Multitrace Fixed Point}
As expected from the structure of the $\beta$-functions, we also discover a fixed point at which only cylcic melonic couplings, that is both single-trace and multitrace cyclic melonic couplings, are nonzero. This fixed point has also been found previously in the complex model. In both cases it has two relevant directions in truncations, see Tables~\ref{tab:CycMult} and \ref{tab:cmmmtFP}. To quartic order the critical exponents of the real model agree with those of the complex one, see Table~\ref{tab:CycMult}. At sixth order, however, the critical exponents obtained here deviate from the complex model. In particular the most relevant eigendirection has an associated critical exponent of 2.23 which deviates by $15\%$ from the most relevant critical exponent in the complex model, see Table~\ref{tab:cmmmtFP}. Furthermore, compared to the complex model, in the real model the fixed point shows a 3-fold degeneracy in a subset of the critical exponents. This is associated to the fact that in this work we are satisfying the upper bound of the $g^{2,i}_{6,1}$ couplings, which was not the case in the analysis in \cite{Eichhorn:2017xhy}. 
Indeed, when evaluating the stability matrix and considering the rows associated to $\beta_{{g}^{2,1}_{6,1}}$, $\beta_{{g}^{3,2}_{6,1}}$ and $\beta_{{g}^{3,3}_{6,1}}$,
one finds that only the diagonal terms contribute non-trivially and take the value $-(4+3\eta)$. It follows that $-(4+3\eta)$ is an eigenvalue of the stability matrix with three-fold degeneracy, cf.~$\theta_{16,17,18}$ in Tab.~\ref{tab:cmmmtFP}\,.\\
Furthermore, 
just as for the cyclic melonic fixed point, there is an enhanced  $O(N)\otimes O(N^2)$ symmetry. 
This explains the existence of a pair of degenerate critical exponents already at quartic level. At the same time, it implies a form of dimensional reduction. Given that $\theta_2$ is rather close to zero, we refrain from speculating whether it could be linked to a multicritical point in the matrix model \cite{Kazakov:1985ea}, as can also be discovered with the FRG, \cite{Eichhorn:2014xaa}.

\begin{widetext}
\begin{table}[H]
	\begin{tabular}{ |c||c|c|c|c|c|} 
		\hline
		scheme   & ${g_{4,1}^{0}}^*$ & ${g_{4,1}^{2,1}}^*$ & ${g_{4,1}^{2,2}}^*$ & ${g_{4,1}^{2,3}}^*$& ${g_{4,2}^{2}}^*$ \\ \hline 
		full  & 0 & -0.29 & 0 &  0 & -0.38  \\
		pert. & 0 & -0.35   & 0 & 0  & -0.38  \\
		$\eta$=0 & 0  & -1  & 0 & 0 & 1.25  \\
		\hline
		\hline
		scheme & $\theta_1$ & $\theta_2$ & $\theta_3$ & $\theta_{4,5}$ & $\eta$\\ \hline
		full & 2.57   &  0.11 & -0.16  & -0.66 & -0.67 \\ 
		pert. & 2.20 & 0.10 & -0.19 & -0.69 & -0.65  \\
		$\eta$=0 &  2 & -1  & -1.5 &  -2 &  0 \\
		\hline
	\end{tabular}\\
	\caption{\label{tab:CycMult}  At quartic order the fixed point possesses two relevant directions. Compared to the complex model, one finds a new irrelevant direction ($\theta_3$) which is associated to the``tetrahedral" interaction. There is good agreement between the complex and the real model at quartic level.}
\end{table}
	\begin{table}[H]
		\begin{tabular}{ |c||c|c|c|c|c|c|c|c|c|c|c|c|c|c|c|} 
			\hline
			sch. & ${g_{4,1}^{0}}^*$ & ${g_{4,1}^{2,1}}^*$  & ${g_{4,1}^{2,(2,3)}}^*$  & ${g_{4,2}^{2}}^*$& ${g^{0,np}_{6,1}}^*$ & ${g^{0,p}_{6,1}}^*$ & ${g^{1,i}_{6,1}}^*$  & ${g^{2,i}_{6,1}}^*$ & ${g^{3,1}_{6,1}}^*$  & ${g^{3,(2,3)}_{6,1}}^*$ & ${g^{3,1}_{6,2}}^*$  & ${g^{3,(2,3)}_{6,2}}^*$ & ${g^{1}_{6,2}}^*$  & ${g^{3}_{6,3}}^*$   \\ \hline 
			full        & 0 & -0.27 & 0 & -0.05  & 0 & 0  & 0 & 0 & -0.13 & 0 &  -0.05 & 0 & 0 & -0.01   \\ 
					pert  & 0 & -0.30 & 0 &  -0.04 & 0 & 0 &  0 & 0 & -0.17 & 0 & -0.05  & 0 & 0 &  -0.01 \\
			$\eta=0$    & 0 & -0.46 & 0 & 0 & 0 & 0  &  0 & 0 & -0.50 & 0 & 0  & 0 & 0 & 0 \\ 
			\hline
		\end{tabular}\\
		\begin{tabular}{ |c||c|c|c|c|c|c|c|c|c|c|c|c|c|c|c|c| }
			\hline
			sch. & $\theta_1$ & $\theta_2$ & $\theta_{3}$ &  $\theta_{4,5}$  & $\theta_{6}$  & $\theta_{7,8}$ & $\theta_{9}$ & $\theta_{10}$  & $\theta_{11,12}$ & $\theta_{13,14}$ & $\theta_{15}$ & $\theta_{16,17,18}$ &$\theta_{19}$ & $\theta_{20,21}$ & $\eta$\\ \hline 
			full & 2.23 & 0.03 & -0.66 &  -1.16  & -1.66 & -1.74 & -2.04 & -2.12  & -2.16 & -2.24 & -2.73 & -2.74 & -3.10 & -3.12 & -0.42\\  
			pert.& 2.03 & 0.03 & -0.67 & -1.17 & -1.67 & -1.76 & -2.05 & -2.14 & -2.17 & -2.26 & -2.77 & -2.76 & -3.10 & -3.14 & -0.41\\
			$\eta=0$  & 2 & 0.53 & -1.5 & -2  & -2.58  & -3 & -2.66 & -3.25 & -3.08 & -3.5 & -3.41 & -4 & -3.94 & -4.08 &-\\ 
			\hline
		\end{tabular}\\
		\caption{\label{tab:cmmmtFP}Fixed-point values and critical exponents in the truncation to sixth order for the  cyclic-melonic multitrace fixed point.}
	\end{table}
\end{widetext}

\section{Candidates for universality classes for 3d quantum gravity}\label{qgcandidate}

\subsection{Isocolored fixed point with tetrahedral interaction}
As a central difference of the rank-3 complex to the real model, the geometric interaction $g_{4,1}^0 T_{abc}T_{ade}T_{fbe}T_{fdc}$ is present. Its dual corresponds to a tetrahedron, inspiring the name ``geometric" interaction. Due to its three-dimensional character, one might expect that a fixed point with a continuum limit that admits a geometric interpretation should feature a finite value for this interaction. Further, in contrast to the cyclic melonic couplings, it does not admit an enhanced symmetry. Accordingly, all three indices are in fact distinct and cannot be summarized to ``super-indices". The corresponding form of dimensional reduction, present at the different cyclic melonic fixed points is therefore not present at this fixed point.\\
	At quartic order of the truncation, a non-vanishing $g_{4,1}^{0,\ast}$ can only be achieved for  $\eta=-0.75$, cf.~Eq.~\eqref{r1g041} and Tab.~\ref{tab:isog4}. In turn, this imposes one condition on the set of couplings $\left\{g_{4,1}^{2,i},g_{4,2}^2\right\}$. To satisfy all fixed-point conditions for this set, $g_{4,1}^{0\, \ast}$ is fixed, even though its own beta function vanishes for any value of $g_{4,1}^{0,\ast}$. At sixth order in the truncation, a nontrivial zero for $g_{4,1}^0$ can be achieved without setting $\eta=-0.75$, cf.~Tab.~\ref{tab:isog6}, as the fixed-point condition also involves $g_{6,1}^{1,i}$, cf.~Eq.~\eqref{r1g041}. Indeed, one can check that the fixed-point values of ${g_{4,1}^{0}}$ and ${g_{6,2}^{1}}$ are related according to 
	\be
	(3/2 +2\eta){g_{4,1\, \ast}^{0}}=1/40 (5-\eta){g_{6,2\, \ast}^{1}},
	\ee
	 as one would expect when solving Eq.~\eqref{r1g041} . In particular, just as in the quartic truncation ${g^0_{4,1\, \ast}} $ can be either positive or negative. \\
	One should, however, handle the interpretation of Tab.~\ref{tab:isog4} and Tab.~\ref{tab:isog6} with care, given that the mechanisms that lead to a non-trivial fixed-point value of the coupling $g_{4,1}^{0}$ are quite different in the two truncations. Further, we highlight that $\eta=-0.75$ is fixed as a fixed-point requirement only in the quartic truncation. This has a strong impact on the critical exponents, as $\eta$ appears everywhere on the diagonal of the stability matrix. The hexic order is the first order in the truncation at which this requirement is absent and the anomalous dimension can adjusts itself dynamically. This results in a significant shift of both $\eta$ and the critical exponents between the quartic and the hexic truncation.\\
Given the considerations in Sec.~\ref{sec:potstab}, this fixed point might feature a stable potential, possibly with a spontaneously broken global symmetry, since $g_{4,1\,\ast}^{2,i}<0$ and $g_{6,1\, \ast}^{3,i}>0$.

	\begin{widetext}
	
\begin{table}[H]
		\begin{tabular}{ |c||c|c|c|c|c|c| } 
			\hline
			scheme   & ${g_{4,1}^{0}}^*$ & ${g_{4,1}^{2,1}}^*$ & ${g_{4,1}^{2,2}}^*$ & ${g_{4,1}^{2,3}}^*$& ${g_{4,2}^{2}}^*$ &\\ \hline 
			full  & $\pm$ 0.07 & -0.04 & -0.04 & -0.04  &  -0.93 &\\
			pert. &$ \pm$ 0.09  & -0.05 & -0.05 & -0.05 & -1.12& \\
			\hline
			\hline
			scheme & $\theta_1$ & $\theta_2$ & $\theta_3$ & $\theta_4$ & $\theta_5$& $\eta$\\ \hline
			full &  2.98 & -0.28 - i 0.22 & -0.28 + i 0.22 & -0.29 & -0.29 &  -0.75\\ 
			pert. & 2.60 & -0.27- i 0.21 & -0.27 + i 0.21 & -0.31 & -0.31 &  -0.75\\
			\hline
		\end{tabular}\\
		\caption{\label{tab:isog4}The isocolored fixed point with tetrahedral interaction $g_{4,1}^0$ features an anomalous dimension of $\eta=-0.75$, as discussed in Sec.~\ref{sec:betas}. It is clear then that for $\eta=0$ no such fixed point exists. The sign of $g_{4,1}^{0\, \ast}$ is not determined by the fixed-point conditions, i.e., there are two fixed points.}
	\end{table}

	\begin{table}[H]
		\begin{tabular}{ |c||c|c|c|c|c|c|c|c|c|c|c|c|c|c| } 
			\hline
			scheme & ${g_{4,1}^{0}}^*$ & ${g_{4,1}^{2,i}}^*$  & ${g_{4,2}^{2}}^*$ & ${g^{0,np}_{6,1}}^*$ & ${g^{0,p}_{6,1}}^*$ & ${g^{1,i}_{6,1}}^*$ & ${g^{2,i}_{6,1}}^*$ & ${g^{3,i}_{6,1}}^*$    &${g^{1}_{6,2}}^*$   \\ \hline 
			full & $\pm$ 0.42 & -0.50  & 3.43  & 0 & 0 &0 & -5.33 & 4.19 & $\pm$ 2.65    \\ 
			pert. & $\pm$ 0.46 & -0.54 & 3.68  & 0 & 0 &0 & -6.33 & 4.96 &  $\pm$ 3.18\\ 
			$\eta=0$  & $\pm$ 0.60  & -0.71 & 5.09  & 0 & 0 &0 & -10.67 & 8.20  & $\pm$ 7.16   \\ 
			\hline
			\hline
			scheme & $\theta_{1,2}$ & $\theta_{3,4}$ & $\theta_{5,6}$ & $\theta_{7,8}$ & $\theta_{9,10}$ & $\theta_{11,12}$ & $\theta_{13,14}$ & $\theta_{15,16,17}$ & $\theta_{18,19}$    \\ \hline 
			full  & 1.35 $\pm$ i 1.56   & 0.13 & -0.02 $\pm$ 5.10 i & -0.08 $\pm$ 5.35 i & -0.08$\pm$ 5.35 i & -0.88 $\pm$ i 1.33 &-1.40 & -1.43 & -2.00 \\ 
			pert. & 1.46 $\pm$ $i$ 1.39  & 0.15 & -0.11 $\pm$ i 4.83 & -0.10 $\pm$ i 5.42  & -0.10 $\pm$ i 5.42  & -0.99 $\pm$ i 1.34 & -1.41 & -1.46 & -2.04\\ 
			$\eta=0$  & 1.95 $\pm$ 0.69 i   & 0.38 & -0.03 $\pm$ 5.96 i & -0.29 $\pm$ 7.06 i   & -0.29 $\pm$ 7.06 i &$-1.74$ & -1.89 $\pm$ 2.72 i & -2.07 & -3 \\ 
			\hline
		\end{tabular}\\[2mm]
		\begin{tabular}{ |c||c|c|c|c|c|c|c|c|c|c|c| } 
			\hline
			scheme     & ${g^{3,i}_{6,2}}^*$ & ${g^{3}_{6,3}}^*$ & \\ \hline 
			full       & -2.26 & 35.27 & \\ 
			pert. & -2.56 & 39.82 &  \\ 
			$\eta=0$      & -4.35 & 67.32 &  \\ 
			\hline
			\hline
			scheme     & $\theta_{20}$ & $\theta_{21}$ & $\eta$\\ \hline 
			  full    & -3.04 &-4.80 & -0.33 \\ 
			pert. & -2.95 & -5.36 & -0.32\\ 
			$\eta=0$    & -5.70+1.57i & -5.70-1.57 i & 0\\ 
			\hline
		\end{tabular}\\
		\caption{\label{tab:isog6} Fixed-point values and critical exponents at sixth order in the truncation for the isocolored fixed point with tetrahedral interaction.}
	\end{table}
\end{widetext}

We tentatively conjecture that this fixed point might be related to the continuum limit to three-dimensional quantum gravity. One way of testing this conjecture would be to directly probe the emergent geometry. Simultaneously, a comparison to the critical exponents calculated on the continuum side, e.g., in the continuum approach to asymptotic safety, can also provide indications whether the universality classes agree. In $d=3$ Euclidean dimensions, the critical exponents from the Einstein Hilbert truncation are found to be $\theta_1\approx 2.5$ and $\theta_2\approx 0.8$ \cite{Biemans:2016rvp} and $\theta_1=1$ is found for a truncation including only the Newton coupling \cite{Falls:2015qga}. No results are available for higher-order truncations.
The discretized Wheeler de Witt equation provides the leading critical exponent as $\theta_1=11/6$ \cite{Hamber:2012zm}.
A rough estimate on the lower bound of our systematic error, arising from the need to truncate, can be given by comparing the critical exponents in the quartic and hexic truncation. We find a deviation $\delta \approx 1.6$ for the leading exponent.
Thus higher precision on the critical exponents is required in order to decide whether these universality classes agree. Nevertheless, we stress that a comparison of the critical exponents is a viable way of learning whether the continuum limit linked to the isocolored fixed point with tetrahedral interaction is three-dimensional quantum gravity.

We find another isocolored fixed point at which the tetrahedral interaction vanishes which is discussed in App.~\ref{app:IMFP}. This fixed point features slightly complex fixed-point values (imaginary parts smaller than one) in the hexic truncation beyond the approximation with $\eta=0$. By mimicking a different choice of regulator shape function, we show that under such a change the fixed-point values become real, indicating the need of larger truncations to decide about the existence of the fixed point.

\section{Conclusions and outlook}\label{sec:conclusions}

In this paper, we have considered the RG flow with tensor size $N$ in real rank-3 tensor models. This setup allows us to test whether a nontrivial, universal large $N$ scaling regime exists. The search for universality is underpinned by an interpretation of the pregeometric RG flow from many to few degrees of freedom, in accordance with the a-theorem.
 A universal fixed-point regime at large $N$ would constitute a candidate for the continuum limit in these models and thus be of interest for quantum gravity. Therefore, the search for universal scaling regimes in this setup is a question of paramount importance. We extend previous work on matrix and tensor models with the FRG framework \cite{Eichhorn:2013isa,Eichhorn:2014xaa,Eichhorn:2017xhy} to the real case and find indications that rank-3 tensor models feature more than one universality class. We find indications that a dimensionally reduced continuum limit reproducing results from matrix models for two-dimensional quantum gravity exists. Further, we find a candidate 
 for a universality class which might correspond to three-dimensional quantum gravity. It features a finite fixed-point value for a tetrahedral (``geometric") interaction and furthermore its symmetry structure prevents the possibility of a dimensional reduction to a matrix model.\\
 
 We also address an intriguing question that arises because of the disassociation of the scaling with $N$ from canonical scaling determined by mass dimension that exists in standard QFTs. The couplings in tensor models, existing in a pregeometric setting, do not feature mass dimensions. Accordingly, it appears that there is an additional freedom in theory space that corresponds to different \emph{choices} of the large $N$ scaling at the origin of theory space (which we refer to as \emph{canonical} scaling although it is not related to mass-dimensions of couplings.) In our study, we address this freedom for the first time. In the FRG setting, it is related to a choice of the scaling of the regulator term with $N$. This fixes the apparent freedom in theory space by determining the large $N$ scaling of the beta functions. In turn, this determines bounds on the canonical dimensions which arise from the requirement of a well-defined large $N$ limit to exist, as discussed in \cite{Eichhorn:2017xhy} and \cite{BenGeloun:2018ekd}. We highlight how insight into the physical interpretation underlying the tensor model can provide an independent way of fixing this freedom. In our case, we argue that  a geometric interpretation of tensor models requires a particular canonical scaling dimension for one of the couplings. This in turn fixes the freedom in the regulator and the scaling of all other couplings. Furthermore, we discover that the universality class of two-dimensional quantum gravity can be encoded in tensor models in a form of dimensional reduction. To recover  
 the leading critical exponent, the scaling of the regulator must be adapted to the two-dimensional case. Therefore, the additional freedom in the choice of canonical scaling dimensions allows to recover matrix-model results from the tensor model. \\
 
The discovery of several distinct fixed points with a potential (in 3d) or established (in the dimensionally reduced 2 d case) interest for quantum gravity within a relatively inexpensive FRG calculation highlights the power of the FRG as a discovery tool for universality classes. As shown in \cite{Eichhorn:2014xaa} for matrix models, similar truncations, when the breaking of symmetry through the regulator is properly accounted for, can even lead to quantitatively robust results.
\\
We employ a truncation of theory space that is not yet sufficient to observe apparent convergence of universal scaling exponents at the quantitative level. On the other hand, we discover two fixed points with very similar properties to those in complex rank-3 models. Both are fixed points which are induced by the interactions common between the real and complex model. Therefore the robust appearance of these scaling regimes in the two different models can be viewed as an indication of the existence of the corresponding two universality classes. In this paper, we link their properties and enhanced symmetry to a dynamical dimensional reduction consisting in the ``merging" of two indices to a superindex. This implies that one of these fixed points in fact reproduces matrix models for quantum gravity. We show explicitly how to use a freedom in our regularization scheme to adjust the canonical scaling dimensions to agree with matrix models. This results in an agreement of the critical exponent for matrix models and tensor models, compatible with the 2d quantum-gravity result within the systematic error of our truncation.

A novel fixed point of particular interest, the isocolored fixed point with tetrahedral interaction, is new to the real model. It features a nonzero interaction,  the dual geometric interpretation of which is the gluing of four triangles to a tetrahedron. This interaction could be associated to one of the two relevant directions that exist in our truncation. This universality class might be a way of going beyond simpler large-$N$ limits in rank three tensor models which do not give rise to an extended three-dimensional geometry. In particular, the symmetry-structure of the interactions prevents a dynamical dimensional reduction as for the other fixed points. Moreover, our results indicate several relevant directions, hinting at a more involved continuum limit. This is also in agreement with evidence from asymptotically safe quantum gravity that 3d quantum gravity features several relevant directions. This provides tentative support for the conjecture, that a background-independent continuum limit, a.k.a. asymptotic safety, exists in 3d quantum gravity, and makes the extension of these results to the rank-4 case a highly interesting endeavour.

Having discovered several potential candidates for a continuum limit in tensor models, a key question is the physical nature of the continuum limit. 
In fact, the continuum limit associated to the corresponding universality classes might not be physically interesting because the emergent phase of random geometries might not resemble an extended 3-dimensional manifold. For instance, the double-scaling limit in rank-3-models explored in \cite{Gurau:2013cbh} appears to lead to a branched-polymer phase. Given the additional relevant directions that the isocolored fixed point features in our truncation, it might provide a way to go beyond the double-scaling limit and tune the model to reach a continuum limit of physical interest.\\
To address the nature of the continuum phase, at least three  strategies appear possible.\\
Firstly, geometric operators should have a mapping into the space of tensor-operators. Then, the expectation value of these operators can be studied once the fixed-point action is known. At present, the construction of such a mapping appears challenging in practice.\\
Secondly, closing the gap to lattice studies of dynamical triangulations will allow to explore the universality class of the continuum limit with the FRG, while studying the large-scale features of the corresponding phase with lattice studies as in \cite{Ambjorn:2007jv}. The basis for such a connection has been laid in {\cite{Benedetti:2011nn}, where a tensor model for Euclidean Dynamical Triangulations has been developed. The application of the FRG to this setting is indeed possible.\\
Finally, one of the key motivations for this work is the assumption that tensor models provide a background-independent way to access the path integral over metric fluctuations. Additionally, topological fluctuations might or might not contribute in the continuum limit. The latter case is the subject of studies into asymptotically safe quantum gravity, where the status of topological fluctuations is in principle open, but where in practice topological fluctuations are presumably suppressed
due to the choice of background topology. As both settings, tensor models and continuum RG studies might provide access to the same path integral, a comparison of critical exponents would provide an answer as to whether they could correspond to
two sides of the same phase transition. Then, the nature of the geometric phase is much easier accessed in the continuum FRG setup from asymptotic safety. In order to pin down such a connection, a quantitatively robust determination of the universality class is necessary on both sides, and requires more elaborate truncations for both settings. Within the current systematic errors on both the continuum and the discrete side which still require substantial reduction, we find that the two sets of critical exponents could be compatible.

In summary, our results, based on a thorough conceptual understanding of background-independent, pregeometric RG flows, provide tentative hints at a universal continuum limit in rank 3 tensor models that could agree with the universality class of the Reuter fixed point in asymptotically safe gravity. This strongly motivates the first studies of rank 4 tensor models with the FRG.

\emph{Acknowledgements}
 We acknowledge inspiring and encouraging discussions with J.~Ben Geloun, S.~Carrozza, R.~Gurau, J.~M.~Pawlowski  and V.~Rivasseau, as well as helpful discussions with the speakers at the workshop ``FRGE for tensor models" in March 2018 at the ITP Heidelberg. A.~E.~and J.~L.~thank R.~Toriumi for the hospitality at OIST during the workshop on holographic tensors.
A.~E., J.~L.~and A.~D.~P.~are supported by the DFG through the Emmy-Noether-program under grant no.~EI/1037-1. T.~K.~was supported by the project PAPIIT IA-103718 from the UNAM during this project and profited from FQXi-MGA-1713 from the Foundational Questions Institute. 

\begin{appendix}
	
	\section{Explicit forms of beta functions}\label{explicit forms of beta functions}
	We list the explicit forms of the beta functions for the rank-3 real model. To show the explicit forms, we define the threshold functions, which rely on a parameter $r$ defined in the regulator, see Eq.~\eqref{equ:rIRsuppressionTerm}: 
	\al{
		{\mathcal I}^1_1 &=\frac{1}{3} \eta\,  N^{r}+\frac{1}{2} N^{r} (r-\eta ),\\
		{\mathcal I}^2_1 &=\frac{1}{4} \eta\,  N^{2 r}+\frac{1}{3} N^{2r} (r-\eta ),\\
		{\mathcal I}^3_1 &=\frac{1}{10} \eta\,  N^{3r}+\frac{1}{8} N^{3r} (r-\eta ),\\
		{\mathcal I}^1_2 &=\frac{1}{4} \eta\,  N^{ r}+\frac{1}{3} N^{ r} (r-\eta ),\\
		{\mathcal I}^2_2 &=\frac{1}{5} \eta\,  N^{2r}+\frac{1}{4} N^{2 r} (r-\eta ),\\
		{\mathcal I}^3_2 &=\frac{1}{12} \eta\,  N^{3 r}+\frac{1}{10} N^{3 r} (r-\eta ),\\
		{\mathcal I}^1_3 &=\frac{1}{5} \eta\,  N^{r}+\frac{1}{4} N^{r} (r-\eta ),\\
		{\mathcal I}^2_3 &=\frac{1}{6} \eta\,  N^{2r}+\frac{1}{5} N^{2r} (r-\eta ),\\
		{\mathcal I}^3_3 &=\frac{1}{14} \eta\,  N^{3r}+\frac{1}{12} N^{3 r} (r-\eta ),
	} 
	where the lower index denotes the order in the $\mathcal{P}^{-1}\mathcal{F}$ expansion and the upper index refers to the number of indices that are integrated over.
	\begin{widetext}
		The anomalous dimension $\eta$ is given by
		\al{
			\eta &= 4\, {\mathcal I}^3_1\,g^{2}_{4,2} N^{[\bar{g}^{2}_{4,2}]}+4\, {\mathcal I}^2_1\, \left(g^{2,1}_{4,1}+g^{2,2}_{4,1}+g^{2,3}_{4,1}\right)N^{[\bar{g}^{2,i}_{4,1}]}+12\, {\mathcal I}^1_1\, \lgCross N^{\ScalingC041}
\label{fulletar}		}
		 The beta functions are 
		\al{
			{\beta}_{\lgCross}=&\left(-[\bar{g}^0_{4,1}]+2\eta\right)\lgCross+ 16\,{\mathcal I}^1_2\,\left(\lgCyclic1+\lgCyclic2+\lgCyclic3\right)\lgCross\, N^{\Scaling241} 
			-{\mathcal I}^2_1 \left(\lgCat1+\lgCat2+\lgCat3\right)\times \nn 
&\times N^{[\bar{g}^{1,i}_{6,1}]-[\bar{g}^0_{4,1}]}
			-{\mathcal I}^3_1\, {g}^1_{6,2} N^{[\bar{g}^1_{6,2}]-[\bar{g}^0_{4,1}]}-6{\mathcal I}^1_1{{g}^{0,p}_{6,1}}N^{[{\bar{g}^{0,p}_{6,1}}]-[\bar{g}^{0}_{4,1}]} \,,
		}
			\al{
			{\beta}_{{g}^{2,i}_{4,1}}=& \left(-[\bar{g}^{2,i}_{4,1}]+2\eta\right){g}^{2,i}_{4,1}
			+8 \mathcal{I}^1_2\,\left({g}_{4,1}^{{0}}\right)^2\, N^{2\ScalingC041-\Scaling241}+16\,\mathcal{I}^1_2\, \left({g}^{2,i}_{4,1}\, \lgCross\right) N^{\ScalingC041}+8\, \mathcal{I}_2^2 \left(\lgCyclicI\right)^2 \times \nn
&\times N^{\Scaling241}-5\,\mathcal{I}^1_1\, \lgCatI\, N^{\Scaling161-\Scaling241}-\mathcal{I}^3_1\,\lgCyclicDisI\, N^{\Scaling362-\Scaling241}-3\,\mathcal{I}^2_1\, \lgCyclicSixI\, N^{\Scaling361-\Scaling241}
			-3\,\I11\,{g}^{0,np}_{6,1}\times \nn
&\times N^{[\bar{g}^{0,np}_{6,1}]-\Scaling241}
			-2\,\I21\,\dgI261\, N^{\Scaling261-\Scaling241}-\I11 {{g}^{0,p}_{6,1}}N^{[{\bar{g}^{0,p}_{6,1}}]-\Scaling241}\,,
		}
		\al{
			{\beta}_{\dgC242}=&\left(-[\bar{g}^2_{4,2}]+2\eta\right)\dgC242
			+16\, \I12 \left(\dg2141\,\dg2241+\dg2141\,\dg2341+\dg2241\,\dg2341\right)N^{2\Scaling241-\ScalingC242}\nn
			&+16\,\I22\, \dgC242 \left(\dg2141+\dg2241+\dg2341\right)\,N^{\Scaling241}+8\, \I32 \left(\dgC242\right)^2\,N^{\ScalingC242}
			-3\,\I31\, \dgC363\, N^{\ScalingC363-\ScalingC242}\nn
			&-2\,\I21\left(\dg3162+\dg3262+\dg3362\right)N^{\Scaling362-\ScalingC242} -\I11\left(\dg2161+\dg2261+\dg2361\right) N^{\Scaling261-\ScalingC242}\nn
			&-6\, \I11\, \dgC162\, N^{\ScalingC162-\ScalingC242}+48\,\I12\,\dgC242\,\dgC041\, N^{\ScalingC041}\,,
		\label{fullbetag242r}} 
		\al{
			{\beta}_{\dg3i61} =&(-\Scaling361+3\eta)\dg3i61+8\, \mathcal{I}^1_2\, \dg1i61 \, \dgC041 N^{\Scaling161+\ScalingC041-\Scaling361}+32\, \mathcal{I}^1_2\, \dg1i61\, \dg2i41 N^{\Scaling161+\Scaling241-\Scaling361}\nn 
			&+48\,\mathcal{I}^1_2\,\dg3i61\,\dgC041\, N^{\ScalingC041}
			-8\,\mathcal{I}^1_2\,\sum_{j,k}^3 \dg2i41\left(\dg2j61+\dg2k61 \right)(\delta_{ij}-1)(\delta_{jk}-1)(\delta_{ik}-1)\, N^{\Scaling241+\Scaling261-\Scaling361}\nn
			& -192\,\mathcal{I}^1_3 \left( \dg2i41\right)^2\, \dgC041\, N^{2\Scaling241+\ScalingC041-\Scaling361}-32\,\mathcal{I}^2_3\, \left( \dg2i41\right)^3 N^{3\Scaling241-\Scaling361}+24\, \mathcal{I}^2_2\, \dg3i61\, \dg2i41 N^{\Scaling241}\,,
		}
		\al{
			{\beta}_{\dg1i61}=&\left(-\Scaling161+3\eta\right)\dg1i61+16\,\mathcal{I}^1_2\, \dg1i61\,\dgC041\, N^{\ScalingC041}+8\,\mathcal{I}^2_2 \,\dg1i61\dg2i41 \, N^{\Scaling241}+24\,\mathcal{I}^1_2\,\dg3i61\,\dgC041 N^{\Scaling361+\ScalingC041-\Scaling161}\nn
			& -8\,\mathcal{I}^1_2\,\sum_{j,k}^3\left(\dg2j61+\dg2k61\right)\dgC041(\delta_{ij}-1)(\delta_{jk}-1)(\delta_{ik}-1)\,N^{\Scaling261+\ScalingC041-\Scaling161}+16\I12 {{g}^{0,p}_{6,1}}{g}^0_{4,1}\times \nn 
&\times N^{[{\bar{g}^{0,p}_{6,1}}]+[\bar{g}^0_{4,1}]-[\bar{g}^{1,i}_{6,1}]}
			+32\I12 {{g}^{0,p}_{6,1}}{g}^{2,i}_{4,1}N^{[{\bar{g}^{0,p}_{6,1}}]+[\bar{g}^{2,i}_{4,1}]-[\bar{g}^{1,i}_{6,1}]} -96 \,\mathcal{I}^1_3 \left(\dg2i41\right)^2\,\dgC041\,N^{2\Scaling241+\ScalingC041-\Scaling161}\nn
			&+96\,\mathcal{I}^1_3\,\sum_{j,k}^3 \dg2i41\left(\dg2j41+\dg2k41\right)\dgC041\,(\delta_{ij}-1)(\delta_{jk}-1)(\delta_{ik}-1)\,N^{2\Scaling241+\ScalingC041-\Scaling161}\,,
		}
		\al{
			{\beta}_{\dgC162}=& \left( -\ScalingC162+3\eta \right)\dgC162+16\,\mathcal{I}^1_2\, \dgC041\left(\dg3162+\dg3262+\dg3362\right)\,N^{\ScalingC041+\Scaling362-\ScalingC162}\nn
			& -4 \, \mathcal{I}_2^1\,\sum_{i,j,k}^3 \dg1i61\left(\dg2j41+\dg2k41\right)(\delta_{ij}-1)(\delta_{jk}-1)(\delta_{ik}-1)\,N^{\Scaling161+\Scaling241-\ScalingC162}+8\,\mathcal{I}^3_2\, \dgC162 \dgC242 N^{\ScalingC242}\nn
			& +8\,\mathcal{I}^2_2\left(\dg1161+\dg1261+\dg1361\right)\dgC242\,N^{\Scaling161+\ScalingC242-\ScalingC162}+16\,\mathcal{I}^2_2\, \dgC162 \left(\dg2141+\dg2241+\dg2341\right)N^{\Scaling241}\nn
		&+24\,\mathcal{I}^1_2 \,\dgC162\, \dgC041\,N^{\ScalingC041}+48\,\I12\,{{g}^{0,p}_{6,1}}{g}^{2}_{4,2}N^{[{\bar{g}^{0,p}_{6,1}}]+[\bar{g}^2_{4,2}]-[\bar{g}^1_{6,2}]}\,, }
		\al{
			{\beta}_{{g}^{0,np}_{6,1}}=& \left(-[\bar{g}^{0,np}_{6,1}]+3\eta \right){g}^{0,np}_{6,1}+ 8\,\mathcal{I}^1_2 \left[ \dg1161(\dg2241+\dg2341)+ \dg1261(\dg2141+\dg2341)+\dg1361(\dg2141+\dg2241)\right]\times \nn
&\times N^{\Scaling161+\Scaling241-[\bar{g}^{0,np}_{6,1}]}
			 +24\,\mathcal{I}^1_2\,{g}^{0,np}_{6,1}\left(\dg2141+\dg2214+\dg2341\right)N^{\Scaling241} -96 \,\mathcal{I}^1_3\,\left(\dgC041\right)^3 N^{3\ScalingC041-[\bar{g}^{0,np}_{6,1}]}\nn 
&-96 \,\mathcal{I}^1_3 \left(\dgC041\right)^2\,\left(\dg2141+\dg2241+\dg2341\right) 
 N^{2\ScalingC041+\Scaling241-[\bar{g}^{0,np}_{6,1}]\,,}
			}
		\al{
			{\beta}_{\dgC363}=&\left(-\ScalingC363+3\eta\right)\dgC363+24\,\mathcal{I}^2_2\,\dgC363\left(\dg2141+\dg2241+\dg2341\right)N^{\Scaling241}+72\,\mathcal{I}^1_2\,\dgC363\,\dgC041\,N^{\ScalingC041}+24\,\mathcal{I}^3_2\,\dgC363\,\dgC242\times\nn 
&\times N^{\ScalingC242}+16\,\mathcal{I}^2_2\,\left(\dg3162+\dg3262+\dg3362\right)\dgC242\,N^{\Scaling362+\ScalingC242-\ScalingC363}+48\,\mathcal{I}^1_2\,\dgC162\,\dgC242 N^{\ScalingC162+\ScalingC242-\ScalingC363}\nn
			& +8\,\mathcal{I}^1_2\,\left(\dg2161+\dg2261+\dg2361\right)\dgC242 N^{\Scaling261+\ScalingC242-\ScalingC363}-192\,\mathcal{I}^1_3\,\dgC242\left(\dg2141\dg2241+\dg2141\dg2341+\dg2241\dg2341\right)\times\nn
&\times N^{2\Scaling241+\ScalingC242-\ScalingC363}-96\,\mathcal{I}^2_3\,\left(\dg2141+\dg2241+\dg2341\right)\left(\dgC242\right)^2 N^{2\ScalingC242 + \Scaling241-\ScalingC363}\nn
&-288\,\mathcal{I}^1_3\,\dgC041\left(\dgC242\right)^2N^{\ScalingC041+2\ScalingC242-\ScalingC363}
			-32\,\mathcal{I}^3_3\,\left(\dgC242\right)^3 N^{3\ScalingC242-\ScalingC363}\,,
		\label{fullbetag363r}}
	
		\al{
			{\beta}_{\dg2i61}=&\left(-\Scaling261+3\eta \right)\dg2i61+96\,\mathcal{I}^1_3\,\sum_{j,k}^3\dg2j41\,\dg2k41\,\dgC041 (\delta_{ij}-1)(\delta_{jk}-1)(\delta_{ik}-1)\,N^{2\Scaling241+\ScalingC041-\Scaling261}\nn
			&+48\,\mathcal{I}^1_3\,\sum_{j,k}^3\left(\dg2j41+\dg2k41\right)\left(\dgC041\right)^2 (\delta_{ij}-1)(\delta_{jk}-1)(\delta_{ik}-1)N^{2\ScalingC041 + \Scaling241-\Scaling261}\nn
			&+96\,\mathcal{I}^1_3\,\sum_{j,k}^3\left(\dg2j41\right)^2\,\dg2k41(\delta_{ij}-1)(\delta_{jk}-1)(\delta_{ik}-1)\,N^{3\Scaling241-\Scaling261}\nn
			& -16\,\mathcal{I}^1_2\,\sum_{j,k}^3 \dg1j61\,\dg2k41(\delta_{ij}-1)(\delta_{jk}-1)(\delta_{ik}-1)\,N^{\Scaling161+\Scaling241-\Scaling261}\nn
			& -8\,\mathcal{I}^1_2\,\sum_{j,k}^3\,\left(\dg1j61+\dg1k61\right)\dgC041(\delta_{ij}-1)(\delta_{jk}-1)(\delta_{ik}-1)N^{\Scaling161+\ScalingC041-\Scaling261}\nn
			& -24\,\mathcal{I}^1_2 \,\sum_{j,k}^3 \dg3j61\,\dg2k41 (\delta_{ij}-1)(\delta_{jk}-1)(\delta_{ik}-1) N^{\Scaling361+\Scaling241-\Scaling261})\nn
			& -12 \,\mathcal{I}^1_2\,\sum_{j,k}^3 \left(\dg2j41+\dg2k41\right){g}^{0,np}_{6,1}(\delta_{ij}-1)(\delta_{jk}-1)(\delta_{ik}-1)N^{\Scaling241+[\bar{g}^{0,np}_{6,1}]-\Scaling261}\nn
			&-4\,\I12\,\sum_{j,k}^3 \left({g}^{2,j}_{4,1}+{g}^{2,k}_{4,1}\right){{g}^{0,p}_{6,1}}(\delta_{ij}-1)(\delta_{jk}-1)(\delta_{ik}-1)N^{[{\bar{g}^{0,p}_{6,1}}]+[\bar{g}^{2,i}_{4,1}]-[\bar{g}^{2,i}_{6,1}]}\nn
			& -4\,\mathcal{I}^2_2\,\sum_{j,k}^3 \left(\dg2j41+\dg2k41\right)\dg2i61 (\delta_{ij}-1)(\delta_{jk}-1)(\delta_{ik}-1)N^{\Scaling241}+40\, \mathcal{I}^1_2 \,\dg2i61\,\dgC041\,N^{\ScalingC041}\nn
			&-192\,\mathcal{I}^1_3\,\dg2141\,\dg2241\,\dg2341\,N^{3\Scaling241-\Scaling261}\,,
		}

		\al{
			{\beta}_{\dg3i62}=&\left(-\Scaling362+3\eta\right)\dg3i62+24 \,\mathcal{I}^2_2\,\dg2i41\dg3i62\,N^{\Scaling241}+56\,\mathcal{I}^1_2\,\dg3i62\,\dgC041\,N^{\ScalingC041}\nn
			&-4\,\mathcal{I}^2_2\,\sum_{j,k}^3\dg3i62\left(\dg2j41+\dg2k41\right)(\delta_{ij}-1)(\delta_{jk}-1)(\delta_{ik}-1)N^{\Scaling241} +8\,\mathcal{I}^{3}_2\,\dg3i62\,\dgC242\,N^{\ScalingC242}\nn
			&-12\,\mathcal{I}^1_2\sum_{j,k}^3 \dg3i61\left(\dg2j41+\dg2k41\right)(\delta_{ij}-1)(\delta_{jk}-1)(\delta_{ik}-1)N^{\Scaling361+\Scaling241-\Scaling362}\nn
			&-4\,\mathcal{I}^2_2\,\sum_{j,k}^3 \dgC242\left(\dg2j61+\dg2k61\right)(\delta_{ij}-1)(\delta_{jk}-1)(\delta_{ik}-1)N^{\ScalingC242+\Scaling261-\Scaling362}\nn
			&-8\,\mathcal{I}^1_2\sum_{j}^3 \dg2j61\,\dg2j41(\delta_{ij}-1)N^{\Scaling261+\Scaling241-\Scaling362}-8\,\mathcal{I}^1_2\sum_{j,k}^3 \dg2i41\left(\dg2j61+\dg2k61\right)(\delta_{ij}-1)(\delta_{jk}-1)(\delta_{ik}-1)\times\nn 
&\times N^{\Scaling241+\Scaling261-\Scaling362}
			+48\,\mathcal{I}^1_3\sum_{j,k}^3 \left(\dg2i41\right)^2\left(\dg2j41+\dg2k41\right)(\delta_{ij}-1)(\delta_{jk}-1)(\delta_{ik}-1)N^{3\Scaling241-\Scaling362}\nn
			&+24\,\mathcal{I}^1_2\,{g}^{0,np}_{6,1}\,\dgC242\,N^{[\bar{g}^{0,np}_{6,1}]+\ScalingC242-\Scaling362}-96\,\mathcal{I}^2_3\,\left(\dg2i41\right)^2\dgC242\,N^{2\Scaling241+\ScalingC242-\Scaling362}-96\,\mathcal{I}^1_3\,\left(\dgC041\right)^2\dgC242\times \nn
&\times N^{2\ScalingC041+\ScalingC242-\Scaling362}
			+8\,\I12\,{{g}^{0,p}_{6,1}}{g}^2_{4,2}N^{[{\bar{g}^{0,p}_{6,1}}]+[\bar{g}^{2}_{4,2}]-[\bar{g}^{3,i}_{6,2}]}+40\,\mathcal{I}^1_2\dg1i61\,\dgC242\,N^{\Scaling161+\ScalingC242-\Scaling362}\nn
&+24\,\mathcal{I}^2_2\,\dg3i61\,\dgC242 N^{[\bar{g}^{3,i}_{6,1}]+\ScalingC242-\Scaling362}
+16\,\mathcal{I}^1_2\,\dgC162\,\dgC041\,N^{\ScalingC162+\ScalingC041-\Scaling362}+32\,\mathcal{I}^1_2\,\dgC162\,\dg2i41\times \nn
&\times N^{\ScalingC162+\Scaling241-\Scaling362} \,,
		}

		\al{
			{{\beta}_{{g}^{0,p}_{6,1}}}=& \left(-[{\bar{g}^{0,p}_{6,1}}]+3\eta\right){{g}^{0,p}_{6,1}}+72\,\I12\,{g}^{0,np}_{6,1}{g}^0_{4,1}N^{[\bar{g}^0_{4,1}]+[\bar{g}^{0,np}_{6,1}]-[{\bar{g}^{0,p}_{6,1}}]}+24\,\I12\,{{g}^{0,p}_{6,1}}{g}^0_{4,1}N^{[\bar{g}^0_{4,1}]}\nn
			&+16\,\I12\,{g}^0_{4,1}\left({g}^{1,1}_{6,1}+{g}^{1,2}_{6,1}+{g}^{1,3}_{6,1}\right)N^{[\bar{g}^0_{4,1}]+[\bar{g}^{1,i}_{6,1}]-[{\bar{g}^{0,p}_{6,1}}]}+24\,\I12\,{{g}^{0,p}_{6,1}}\left({g}^{2,1}_{4,1}+{g}^{2,2}_{4,1}+{g}^{2,3}_{4,1}\right)N^{[\bar{g}^{2,i}_{4,1}]}\,.
		}
	\end{widetext}
	
	\section{Isocolored Melonic Fixed Point}\label{app:IMFP}
	This fixed point is an isocolored melonic fixed point at which the tetrahedral interaction vanishes. In the hexic truncation, it becomes complex, calling its existence into question.
	\begin{widetext}
		\begin{table}[H]
			\begin{tabular}{ |c||c|c|c|c|c|c| } 
				\hline
				scheme & ${g_{4,1}^{0}}^*$  & ${g_{4,1}^{2,1}}^*$ & ${g_{4,1}^{2,2}}^*$ & ${g_{4,1}^{2,3}}^*$ & ${g_{4,2}^{2}}^*$ &\\ \hline 
				full & 0  & -0.31 & -0.31 & -0.31  & 1.45 & \\
				pert& 0   &-0.35 & -0.35 & -0.35& 1.52   &\\
				$\eta=0$ & 0  & -0.16 & -0.16 & -0.16 & -0.69 - i 0.15 & \\
				\hline
				\hline
				scheme & $\theta_1$ & $\theta_2$ & $\theta_3$ & $\theta_4$ & $\theta_5$& $\eta$\\ \hline 
				full  & 1.09 + i 0.87& 1.09 - i 0.87& 0.71 &  0.71 & -0.21&-0.64 \\ 
				 pert& 1.04+ i 0.78 & 1.04 - i 0.78 & 0.70 & 0.70 & -0.2 &  -0.65\\
				$\eta=0$ & 2 & -0.23 i & -0.53 & -0.53 & -1.5&\\ 
				\hline
			\end{tabular}\\
			\caption{The isocolored fixed point features four relevant directions in the truncation to quartic order in the tensors.}
		\end{table}

		\begin{table}[H]
			\begin{tabular}{ |c||c|c|c|c|c|c|c|c|c|c|c|c|c| } 
				\hline
				scheme & ${g_{4,1}^{0}}^*$ & ${g_{4,1}^{2,i}}^*$  & ${g_{4,2}^{2}}^*$ & ${g^{0,NP}_{6,1}}^*$ & ${g^{0,P}_{6,1}}^*$ & ${g^{1,i}_{6,1}}^*$ & ${g^{2,i}_{6,1}}^*$ & ${g^{3,i}_{6,1}}^*$      \\ \hline 
				full & 0 & -0.18 + i 0.04  & 0.36 -i 0.40   & 0 & 0 &0 &0 & -0.03 + i 0.03    \\ 
				 pert. & 0 & -0.27 + i 0.01 & 0.45 - i 0.06 & 0 & 0 & 0 & 0 & -0.11 + i 0.01\\ 
				 $\eta=0$  & 0 & -0.34 &  -0.31 & 0 & 0 & 0 & 0 & -0.13   \\ 
				\hline
				\hline
				scheme & $\theta_{1}$ & $\theta_{2}$ & $\theta_{3,4}$ & $\theta_{5}$ & $\theta_{6}$ & $\theta_{7,8}$ & $\theta_{9}$ & $\theta_{10,11,12}$   \\ \hline 
				full.& 1.66 - i 0.12 & 1.42 + i 1.32  & 0.76 - i 0.13  & -0.25 + i 0.10 & -0.53 - i 0.02 & -1.13 + i 0.15 & -1.19 - i 0.75 & -1.23 + i 0.06  \\  
				pert. & 1.51 - i 0.18 & 1.31 + i 1.32 & 0.76 - i 0.13 & -0.27 + i 0.09 & -0.52 - i 0.02  & -1.15 + i 0.14 & -1.14 - i 0.76 & -1.24 + i 0.06  \\ 
				  $\eta=0$  & 4.83 &  2.39 & 1.46 & -0.18 & -1.33 & -2.17 & -1.5  & -2.64 \\ 
				\hline
			\end{tabular}\\[2mm]
			\begin{tabular}{ |c||c|c|c|c|c|c|c| } 
				\hline
				scheme     & ${g^{1}_{6,2}}^*$  & ${g^{3,i}_{6,2}}^*$ & ${g^{3}_{6,3}}^*$ & & &   \\ \hline 
				full       & 0  &-0.16- i 0.15  & 2.00 + i 0.59 & & & \\ 
				 pert. & 0  & -0.37 - i 0.05 & 4.19 + i 0.27 & & &\\ 
				 $\eta=0$      & 0 & -2.57 & 19.71 & & & \\ 
				\hline
				\hline
				scheme     & $\theta_{13,14,15}$ & $\theta_{16,17}$  & $\theta_{18,19}$ &$\theta_{20}$ &$\theta_{21}$& $\eta$ \\ \hline 
				 full  & -1.32 - i 0.03 & -1.46+ i 0.23 & -1.60 - i 0.21 &-1.65 - i 0.08 & -2.44+ i 0.38 &-0.62 - i 0.05\\
				 pert. & -1.33 - i 0.03 &  -1.47 + i 0.22 &  -1.59 - i 0.21 & -1.67 - i 0.75& -2.47- i 0.36 & -0.62 - i 0.05\\ 
				 $\eta=0$    & -2.82 &  -3 &  -3.26 & -3.71+ i 0.75& -3.71- i 0.75 & 0\\ 
				\hline
			\end{tabular}\\
			\caption{Fixed-point values and critical exponents at sixth order in the truncation for the isocolored melonic fixed point. It is possible to adiabatically connect the $\eta=0$-fixed point to the non-perturbative one by introducing a small parameter $\epsilon$ in front of the anomalous dimension.}
		\end{table}
	\end{widetext}
	From the structure of the beta functions, it is clear that one can consistently set $g_{4,1}^0=0, \, g_{6,2}^1=0,\,g_{6,1}^{1,i}=0$. Thus, a fixed point with the same structure as in the complex rank 3 model, i.e., without the geometric interaction, should also exist at sixth order in the truncation. It remains to check whether the fixed-point values are complex. By exploiting the expected symmetry-structure of the fixed point, we can reduce the set of independent beta functions to six, namely $\beta_{g_{4,1}^{2,1}},\, \beta_{g_{2,4}^2}, \, \beta_{g_{6,1}^{3,1}}, \, \beta_{g_{6,1}^{2,1}}, \, \beta_{g_{6,2}^{3,1}}$ and $\beta_{g_{6,3}^3}$.  It is also worth pointing out that the fixed-point results are regulator-dependent. Therefore choosing a different regulator might indeed alter the numerical fixed-point values leading to real fixed points. Indeed, it is possible to modify our original  regulator by a rescaling with an overall prefactor $x$ such as
		\begin{eqnarray}
		R_N&& (\left\{a_i\right\},\left\{b_i\right\}) =x\, Z_N \,\delta_{a_1 b_1}\delta_{a_2 b_2} \delta_{a_3 b_3}\nonumber\\&& \times\left(\frac{N}{a_1+a_2+a_3}-1\right) \theta \left(\frac{N}{a_1+a_2+a_3}-1\right)\,.
		\label{RescaleRegulator}
		\end{eqnarray} 	
		For $\mathcal{O}(1)$ values of $x$, e.g., $x=2.2$, it is already possible to obtain real fixed-point values.  A more realistic way to account for changes in the regulator is achieved by modifying the threshold functions derived from the regulator rather than rescaling the regulator itself. A possible choice would be for instance the following
		\begin{equation}
		\mathcal{I}^i_\alpha=c_1(i) \eta N^i+c_2(i) N^i(x(\alpha)-\eta),
		\end{equation}
		where $c_1(i)$ and $c_2(i)$ are two constants that depend on the number of indices that are integrated over and $x(\alpha)$ is precisely the factor that captures the re-parametrization of the regulator and depends now, in contrast to the case in Eq.~\eqref{RescaleRegulator}, on the order of the $\mathcal{P}^{-1}\mathcal{F}$-expansion $\alpha$. \\
		In this discussion one should keep in mind that it is not clear whether complex fixed points are actually be discarded from stability arguments of the potential. Given that we work in a background-independent setting, there is no metric and correspondingly no signature, so the action might even be allowed to be complex.
			
	\begin{widetext}
		\begin{center}
			\begin{figure}
				\includegraphics[scale=.3]{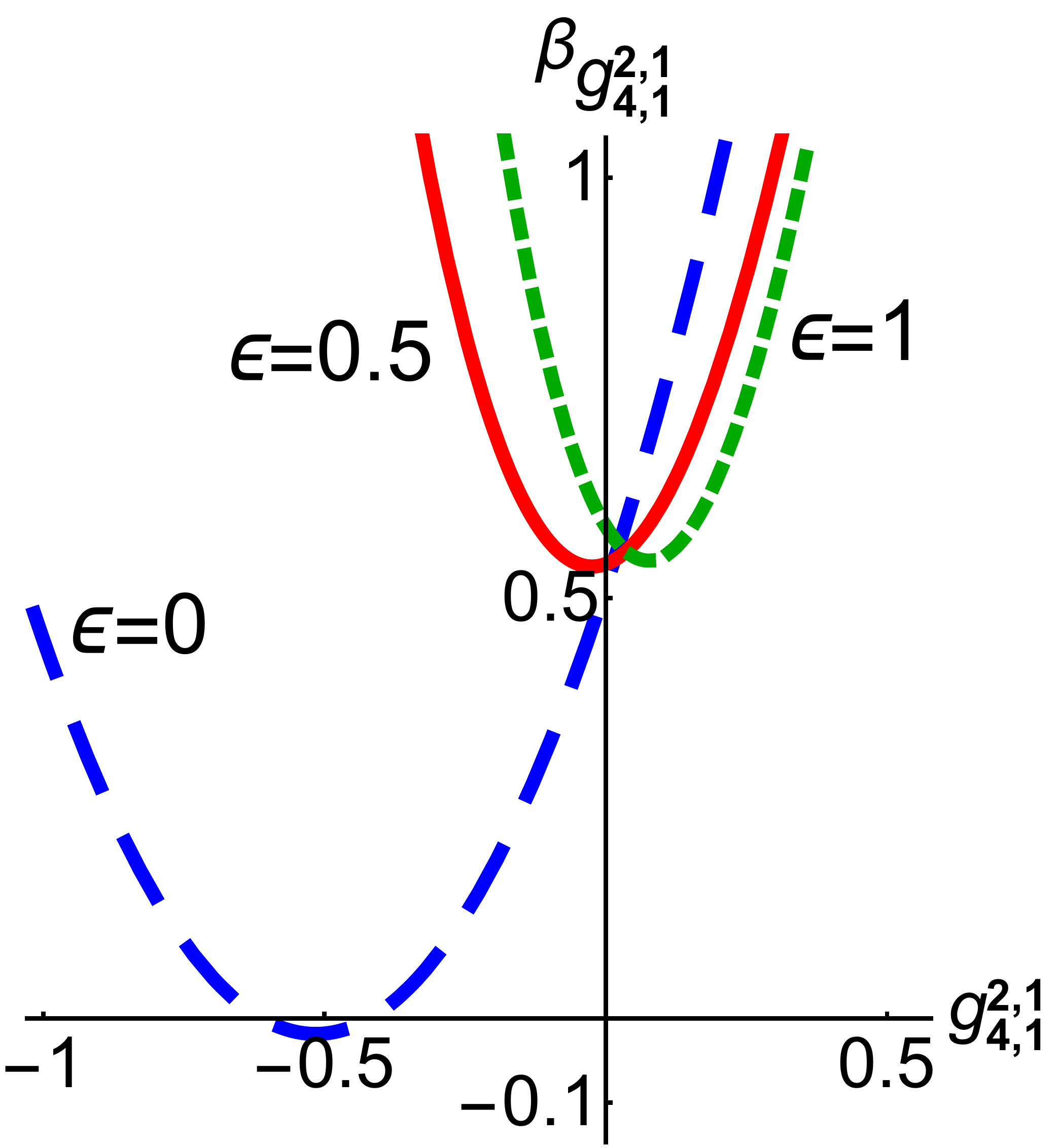}
				\caption{ We multiply the anomalous dimension by a small coefficient $\epsilon$. In this case $\epsilon=1$ refers to the full non-perturbative sixth-order truncation while $\epsilon=0$ implies $\eta_N=0$. In order to check whether a fixed point survives after including the non-perturbative anomalous dimension, we adiabatically vary $\epsilon$ from 1 to 0. The above figure schematically shows how the fixed point shifts from the real axis at $\epsilon=0$ into the complex plane at $\epsilon=1$. }
				\label{IsoMelonic}
			\end{figure}
		\end{center}

			\begin{table}[H]
		\begin{tabular}{ |c||c|c|c|c|c|c|c|c|c|c|c|c|c| } 
			\hline
			scheme & ${g_{4,1}^{0}}^*$ & ${g_{4,1}^{2,i}}^*$  & ${g_{4,2}^{2}}^*$ & ${g^{0,NP}_{6,1}}^*$ & ${g^{0,P}_{6,1}}^*$ & ${g^{1,i}_{6,1}}^*$ & ${g^{2,i}_{6,1}}^*$ & ${g^{3,i}_{6,1}}^*$      \\ \hline 
			full & 0 & -0.11  & 0.15  & 0 & 0 & 0 & 0 & -0.02    \\ 
			 pert. & 0 & -0.12  & 0.14 & 0 & 0 & 0 & 0 & -0.02		\\ 
			$\eta=0$  & 0 & -0.15 & -0.14  & 0 & 0 & 0  & 0 & -0.03    \\ 
			\hline
			\hline
			scheme & $\theta_{1}$ & $\theta_{2}$ & $\theta_{3,4}$ & $\theta_{5}$ & $\theta_{6}$ & $\theta_{7}$ & $\theta_{8,9}$ & $\theta_{10,11,12}$   \\ \hline 
			 full.& 1.89 - i 0.87  & 1.89 + i 0.87  & 1.06  &  -0.32 & -0.37  & -0.67 & -1.675  & -1.684  \\  
			 pert. & 1.87 - i 0.83  & 1.87 + i 0.83 & 1.06 & -0.32 & -0.46 & -0.67 & -1.68 & -1.70  \\ 
			 $\eta=0$  & 4.82 & 2.39 & 1.46 & -0.18 & -1.33 & -1.5 & -2.17 & -2.64  \\ 
			\hline
		\end{tabular}\\[2mm]
		\begin{tabular}{ |c||c|c|c|c|c|c|c| } 
			\hline
			scheme     & ${g^{1}_{6,2}}^*$  & ${g^{3,i}_{6,2}}^*$ & ${g^{3}_{6,3}}^*$ & & &   \\ \hline 
			full       & 0  & -0.09 & -0.90 & & & \\ 
			 pert. & 0  & -0.11 & 1.09 & & &\\ 
			 $\eta=0$      & 0  & -0.53  & 4.07 & & & \\ 
			\hline
			\hline
			scheme     & $\theta_{13,14,15}$ & $\theta_{16,17}$  & $\theta_{18,19}$ &$\theta_{20}$ &$\theta_{21}$& $\eta$ \\ \hline 
			  full  & -1.72 & -1.75 & -1.99 & -2.63 + i 0.08 & -2.63 - 0.08 & -0.42  \\
			 pert. & -1.73 &  -1.76 &  -2.01 & -2.64 + i 0.13 & -2.64 - i 0.13 & -0.41\\ 
			 $\eta=0$    & -2.82 & -3 & -3.26 & -3.71 + i 0.74 & -3.71 - i 0.74 & 0\\ 
			\hline
		\end{tabular}\\
		\caption{Fixed-point values and critical exponents at sixth order in the truncation for the isocolored melonic fixed point for a different regulator. It is possible to adiabatically connect the $\eta=0$-fixed point to the non-perturbative one by introducing a small parameter $\epsilon$ in front of the anomalous dimension.}
	\end{table}

	\end{widetext}
\end{appendix}

\bibliography{refsTensor}
\end{document}